\documentclass[a4paper,11pt]{article}
\pdfoutput=1 % if your are submitting a pdflatex (i.e. if you have
\usepackage{jheppub} % for details on the use of the package, please

\usepackage[T1]{fontenc} % if needed

\newcommand{\SKIP}[1]{}

\title{\boldmath The method of kinematic limits in high-energy physics}

\author[a]{A. V. Bobrov}

\affiliation[a]{BINP,\\ 11 Acad. Lavrentieva Pr. Novosibirs 630090, Russia}

\emailAdd{A.V.Bobrov@inp.nsk.su }

\usepackage[english]{babel} %

\abstract{

  This paper proposes a general approach for calculating kinematic
  limits attained when Lorentz invariants, which are analogous
  to Cayley-Menger determinants for Minkowski space, vanish.
  This approach can be applied to a wide range of processes
  in particle physics.
In particular, this may be relevant for reactions involving lost particles, which can be lost due to both detector inefficiency and the small cross section of particle interactions with detector materials, such as neutrinos. Furthermore, kinematic limits can be used to suppress background.
  
   }

\begin{document} 
\maketitle
\flushbottom

\section{Introduction}
\label{sec:int}

Experimental particle physics underwent a period of rapid development from the 1970s to the 1990s. During this time, a large number of fundamental elementary particles were discovered. One of which was the tau lepton. With the development of experimental methods at the turn of the century, two experiments (Belle and $BABAR$) emerged in which $\tau^{+}\tau^{-}$ pairs were produced in large quantities in electron-positron collisions.
Since all known decays of the $\tau$-lepton involve  neutrinos in the final state, which are not reconstructed in the detector, the study of such processes presents experimentalists with a specific set of questions related to particle kinematics.

 It should be noted that the mathematical aspects of this subject matter were partially understood as early as the 1960s and 1970s \cite{1}. However, the problem was not explicitly formulated, nor was its connection to research in geometry existing at the time fully recognized. An experimental study of $\tau^{+}\tau^{-}$ pair decay processes, combined with a review of prior work on elementary particle kinematics and $\tau$-lepton properties \cite{c}, enabled us to understand the specific challenges involved in analyzing processes with incomplete reconstruction.
  These investigations resulted in the formalization of the kinematic limits method. Furthermore, we address an existing gap in the literature and demonstrated a direct link between the proposed method and geometric studies dating back to the 1930s \cite{a}.

Let's begin our consideration with the abstract process of transformation
of the initial state particles in into the final state particles.
The equation expressing the law of conservation of energy-momentum has the form
$\Sigma_{i=1}^{i=N}sign_{i}p_{i}^{\alpha}=0$, where $p_{i}^{\alpha}$
-- four-vectors of energy-momentum of the initial and final particles,
$sign_{i}=1$ for initial, $sign_{i}=-1$ for final particles.
If N particles (here $p^{2}_{i}=m^{2}_{i}$ is implied) participate in a reaction, then the total number of independent Lorentz invariants is 3N-10.
It is assumed that the particles are not polarized, or their polarization is not measured.
The basic idea is to construct a coordinate system based directly on the four-particle momenta involved in the process.
In this approach, non-invariant degrees of freedom do not exist in principle.
Such attempts are regularly made in particle physics, when the
transition to the rest frame of one particls occurs, and the x-axis is directed along the motion of the second particle
in this coordinate system. Instead, it is proposed to write the equations in a covariant form,
where there is no external coordinate system, and the coordinates
are constructed on the four-vectors of the particles involved in the reaction.
For this specific form of coordinate notation, conditions of the type $(p_{i}\pm p_{j})^2=s_{ij}$ (here $s_{ij}$ are independent
invariants) will be easy to implement. These are essentially the poles of the propagators of the intermediate particles
in Feynman diagrams. An example of such a parameterization is Mandelstam variables.

\section{Special coordinate system}
\label{sec:geo}

We will construct a coordinate system on four-vectors $a, b, c, d$.
If they are linearly independent, then $\epsilon^{\alpha\beta\gamma\rho}a_{\alpha}b_{\beta}c_{\gamma}d_{\rho}\neq0$, so
any four-vector $v$ can be written as:

 \begin{eqnarray}
   v^{\mu}(A,B,C,D)=\frac{\displaystyle[Ab_{\theta}c_{\delta}d_{\psi}+Bc_{\theta}a_{\delta}d_{\psi}+Ca_{\theta}b_{\delta}d_{\psi}+Db_{\theta}a_{\delta}c_{\psi}]\epsilon^{\mu\theta\delta\psi}}{\displaystyle \epsilon^{\alpha\beta\gamma\rho}a_{\alpha}b_{\beta}c_{\gamma}d_{\rho}},
   \label{f:4}
  \end{eqnarray}

where $A=(va), ~ B=(vb), ~ C=(vc), ~D=(vd)$ are the scalar products
of the four-vector $v$ with the basis vectors or coordinates of the four-vector in the new basis $a, b, c, d$.

We define a pair of four-vectors $v^{\mu}_{1}(A_{1},B_{1},C_{1},D_{1})$ and
$v^{\mu}_{2}(A_{2},B_{2},C_{2},D_{2})$.
Considering that the convolution of the Levi-Civita symbols with the Kronecker delta symbol is equal to:

\begin{eqnarray}
\delta_{\mu_{1}}^{\mu_{2}}\epsilon^{\mu_{1}\theta_{1}\delta_{1}\psi_{1}}\epsilon_{\mu_{2}\theta_{2}\delta_{2}\psi_{2}}=\nonumber \\
-\delta^{\theta_{1}}_{\theta_{2}}\delta^{\delta_{1}}_{\delta_{2}}\delta^{\psi_{1}}_{\psi_{2}}+\delta^{\theta_{1}}_{\delta_{2}}\delta^{\delta_{1}}_{\theta_{2}}\delta^{\psi_{1}}_{\psi_{2}}-\delta^{\theta_{1}}_{\delta_{2}}\delta^{\delta_{1}}_{\psi_{2}}\delta^{\psi_{1}}_{\theta_{2}}-\delta^{\theta_{1}}_{\psi_{2}}\delta^{\delta_{1}}_{\theta_{2}}\delta^{\psi_{1}}_{\delta_{2}}+\delta^{\theta_{1}}_{\psi_{2}}\delta^{\delta_{1}}_{\delta_{2}}\delta^{\psi_{1}}_{\theta_{2}}+\delta^{\theta_{1}}_{\theta_{2}}\delta^{\delta_{1}}_{\psi_{2}}\delta^{\psi_{1}}_{\delta_{2}},
\end{eqnarray}

we obtain the following expression for the scalar product:

\begin{eqnarray}
(v_{1}v_{2})=\frac{\displaystyle A_{1}A_{2}[-b^{2}c^{2}d^{2}+b^{2}(cd)^{2}-2(bc)(cd)(bd)+c^{2}(ad)^{2}+d^{2}(bc)^{2}]}{[\epsilon^{\alpha\beta\gamma\rho}a_{\alpha}b_{\beta}c_{\gamma}d_{\rho}]^{2}}+\nonumber \\
+\frac{\displaystyle [A_{1}B_{2}+B_{1}A_{2}][-(bc)(ac)d^{2}+(ab)c^{2}d^{2}-(ab)(cd)^{2}-(ad)(bd)c^{2}+(ac)(bd)(cd)+(ad)(bc)(cd)]}{[\epsilon^{\alpha\beta\gamma\rho}a_{\alpha}b_{\beta}c_{\gamma}d_{\rho}]^{2}}+\nonumber \\
+\frac{\displaystyle B_{1}B_{2}[-a^{2}c^{2}d^{2}+c^{2}(ad)^{2}-2(ac)(ad)(cd)+d^{2}(ac)^{2}+a^{2}(cd)^{2}]}{\displaystyle[\epsilon^{\alpha\beta\gamma\rho}a_{\alpha}b_{\beta}c_{\gamma}d_{\rho}]^{2}}+\nonumber \\
+\frac{\displaystyle [C_{1}B_{2}+B_{1}C_{2}][-(ab)(ac)d^{2}+(bc)a^{2}d^{2}-(bc)(ad)^{2}-(cd)(bd)a^{2}+(ac)(bd)(ad)+(cd)(ab)(ad)]}{\displaystyle [\epsilon^{\alpha\beta\gamma\rho}a_{\alpha}b_{\beta}c_{\gamma}d_{\rho}]^{2}}+\nonumber \\
+\frac{\displaystyle[C_{1}A_{2}+A_{1}C_{2}][-(bc)(ab)d^{2}+(ac)b^{2}d^{2}-(ac)(bd)^{2}-(ad)(cd)b^{2}+(ab)(cd)(bd)+(ad)(bc)(bd)]}{\displaystyle [\epsilon^{\alpha\beta\gamma\rho}a_{\alpha}b_{\beta}c_{\gamma}d_{\rho}]^{2}} +\nonumber \\
+\frac{\displaystyle C_{1}C_{2}[-a^{2}b^{2}d^{2}+a^{2}(bd)^{2}-2(ab)(bd)(ad)+b^{2}(ad)^{2}+d^{2}(ab)^{2}]}{\displaystyle [\epsilon^{\alpha\beta\gamma\rho}a_{\alpha}b_{\beta}c_{\gamma}d_{\rho}]^{2}} +\nonumber \\
+\frac{\displaystyle[A_{1}D_{2}+A_{2}D_{1}][-(cd)(ac)b^{2}+(ad)c^{2}b^{2}-(ad)(cb)^{2}-(ab)(bd)c^{2}+(ac)(bd)(bc)+(ab)(cd)(bc)]}{\displaystyle [\epsilon^{\alpha\beta\gamma\rho}a_{\alpha}b_{\beta}c_{\gamma}d_{\rho}]^{2}} +\nonumber \\
+\frac{\displaystyle D_{1}D_{2}[-a^{2}b^{2}c^{2}+a^{2}(bc)^{2}-2(ab)(bc)(ac)+b^{2}(ac)^{2}+c^{2}(ab)^{2}]}{\displaystyle [\epsilon^{\alpha\beta\gamma\rho}a_{\alpha}b_{\beta}c_{\gamma}d_{\rho}]^{2}} +\nonumber \\
+\frac{\displaystyle[C_{1}D_{2}+D_{1}C_{2}][-(ad)(ac)b^{2}+(cd)a^{2}b^{2}-(cd)(ab)^{2}-(cb)(bd)a^{2}+(ac)(bd)(ba)+(cb)(ad)(ab)]}{\displaystyle [\epsilon^{\alpha\beta\gamma\rho}a_{\alpha}b_{\beta}c_{\gamma}d_{\rho}]^{2}} +\nonumber \\
+\frac{\displaystyle[D_{1}B_{2}+B_{1}D_{2}][-(cd)(bc)a^{2}+(bd)c^{2}a^{2}-(bd)(ac)^{2}-(ab)(ad)c^{2}+(bc)(ad)(ac)+(ab)(cd)(ac)]}{\displaystyle [\epsilon^{\alpha\beta\gamma\rho}a_{\alpha}b_{\beta}c_{\gamma}d_{\rho}]^{2}}.\nonumber \\
\label{f:metric}
\end{eqnarray}

If there are three linearly independent four-vectors
$a,b,c$, the coordinates of the fourth four-vector $d^{\rho}$ can be expressed as $\epsilon^{\rho\alpha\beta\gamma}a_{\alpha}b_{\beta}c_{\gamma}$.
Substituting
$d^{\rho}=\epsilon^{\rho\alpha\beta\gamma}a_{\alpha}b_{\beta}c_{\gamma}$ into the relation (\ref{f:metric}) yields:

\begin{eqnarray}
(v_{1}v_{2})=\frac{\displaystyle A_{1}A_{2}[-b^{2}c^{2}+(bc)^{2}]+[A_{1}B_{2}+B_{1}A_{2}][-(bc)(ac)+(ab)c^{2}]}{\epsilon^{\alpha\beta\gamma\rho}a_{\alpha}b_{\beta}c_{\gamma}\epsilon_{\alpha_{1}\beta_{1}\gamma_{1}\rho}a^{\alpha_{1}}b^{\beta_{1}}c^{\gamma_{1}}}+\nonumber \\
+\frac{\displaystyle B_{1}B_{2}[-a^{2}c^{2}+(ac)^{2}]+[C_{1}B_{2}+B_{1}C_{2}][-(ba)(ca)+(cb)a^{2}]}{\epsilon^{\alpha\beta\gamma\rho}a_{\alpha}b_{\beta}c_{\gamma}\epsilon_{\alpha_{1}\beta_{1}\gamma_{1}\rho}a^{\alpha_{1}}b^{\beta_{1}}c^{\gamma_{1}}}+\nonumber \\
+\frac{\displaystyle [C_{1}A_{2}+A_{1}C_{2}][-(bc)(ab)+(ac)b^{2}]+C_{1}C_{2}[-a^{2}b^{2}+(ab)^{2}]+D_{1}D_{2}}{\displaystyle \epsilon^{\alpha\beta\gamma\rho}a_{\alpha}b_{\beta}c_{\gamma}\epsilon_{\alpha_{1}\beta_{1}\gamma_{1}\rho}a^{\alpha_{1}}b^{\beta_{1}}c^{\gamma_{1}}}. 
\label{f:metric20}
\end{eqnarray}

A similar consideration is possible for three-dimensional space-time with signature $(+;-;-)$. For three-vectors $a,b,c$ such that $\epsilon^{\alpha\beta\gamma}a_{\alpha}b_{\beta}c_{\gamma}\neq0$, any three-vector $u$ can be represented as:

\begin{eqnarray}
u^{\mu}(A,B,C)=\frac{\displaystyle[Ab_{\theta}c_{\delta}+Bc_{\theta}a_{\delta}+Ca_{\theta}b_{\delta}]\epsilon^{\mu\theta\delta}}{\displaystyle \epsilon^{\alpha\beta\gamma}a_{\alpha}b_{\beta}c_{\gamma}}.  
  \label{f:3}
\end{eqnarray}

In this case, the convolution of the Levi-Civita symbols with the Kronecker delta symbol is:

\begin{eqnarray}
\delta_{\mu_{1}}^{\mu_{2}}\epsilon^{\mu_{1}\theta_{1}\delta_{1}}\epsilon_{\mu_{2}\theta_{2}\delta_{2}}=\delta^{\theta_{1}}_{\theta_{2}}\delta^{\delta_{1}}_{\delta_{2}}-\delta^{\theta_{1}}_{\delta_{2}}\delta^{\delta_{1}}_{\theta_{2}},
\end{eqnarray}

and the scalar product of two three-vectors $u^{\mu}_{1}(A_{1},B_{1},C_{1})$
and $u^{\mu}_{2}(A_{2},B_{2},C_{2})$ is given by the expression:

\begin{eqnarray}
(u_{1}u_{2})=\frac{\displaystyle A_{1}A_{2}[b^{2}c^{2}-(bc)^{2}]+[A_{1}B_{2}+B_{1}A_{2}][(bc)(ac)-(ab)c^{2}]+B_{1}B_{2}[a^{2}c^{2}-(ac)^{2}]}{[\epsilon^{\alpha\beta\gamma}a_{\alpha}b_{\beta}c_{\gamma}]^{2}}+\nonumber \\
+\frac{\displaystyle[C_{1}A_{2}+A_{1}C_{2}][(bc)(ab)-(ac)b^{2}]+ [C_{1}B_{2}+B_{1}C_{2}][(ab)(ac)-(bc)a^{2}]+C_{1}C_{2}[a^{2}b^{2}-(ab)^{2}]}{\displaystyle[\epsilon^{\alpha\beta\gamma}a_{\alpha}b_{\beta}c_{\gamma}]^{2}}.\nonumber \\
\label{f:metric1}
\end{eqnarray}

Similarly, substituting $c^{\rho}=\epsilon^{\rho\alpha\beta}a_{\alpha}b_{\beta}$ into the expression for the three-dimensional case (\ref{f:metric1}) yields:

\begin{eqnarray}
  (u_{1}u_{2})=\frac{\displaystyle A_{1}A_{2}b^{2}-[A_{1}B_{2}+B_{1}A_{2}](ab)+B_{1}B_{2}a^{2} +C_{1}C_{2}}{\epsilon^{\alpha\beta\gamma}a_{\alpha}b_{\beta}\epsilon_{\alpha_{1}\beta_{1}\gamma}a^{\alpha_{1}}b^{\beta_{1}}}=\nonumber \\
  =\frac{\displaystyle A_{1}A_{2}b^{2}-[A_{1}B_{2}+B_{1}A_{2}](ab)+B_{1}B_{2}a^{2} +C_{1}C_{2}}{a^{2}b^{2}-(ab)^{2}}. 
  \label{f:metric10}
\end{eqnarray}

\section{The idea of kinematic limits}
\label{sec:kin}

A kinematic limit
can be obtained from the process of dividing an unreconstructed system into two subsystems
or two particles. This process can be described by
the condition $q^{2}=\mu^{2}$, where
$q$ is the difference between the four-momenta of the two subsystems, $\mu^{2}$ is
a parameter calculated from a complete set of independent invariants.
The scalar product of two four-vectors $(v_{1}v_{2})$ (\ref{f:metric20}) does not contain terms proportional to $D_{i}$ to the first power, because $D_{i}$ are multiplied by $\epsilon^{\rho\alpha\beta\gamma}a_{\alpha}b_{\beta}c_{\gamma}h_{\rho}$ with $h_{\rho}$ equal to $a$ or $b$ or $c$, which is zero. In turn, $D_{1}D_{2}$
is multiplied by
$\epsilon^{\mu\theta_{1}\delta_{1}\psi_{1}}\epsilon_{\mu\theta_{2}\delta_{2}\psi_{2}}a_{\theta_{1}}b_{\delta_{1}}c_{\psi_{1}}a^{\theta_{2}}b^{\delta_{2}}c^{\psi_{2}}$, which is not equal to 0.
Calculating $q^{2}$ in coordinates based on four-vectors
$a,b,c$, if $q_{\rho}\epsilon^{\rho\alpha\beta\gamma}a_{\alpha}b_{\beta}c_{\gamma}=D$, we obtain $q^{2}=\alpha+\beta D^{2}$,
where $\alpha,\beta$ are independent of $D$ and $\beta\neq0$.
the four-vector square $q^2$ must be expressed through as an independent set of Lorentz invariants $(qa),(qb),(qc)$.
If two scalar products of the four-vector $q$ with other four-vectors are known, three-dimensional relations can be used.
Taking the given value $q^2=\mu^2$,  $D^2$ is given by $D^2= (\mu^2 - \alpha)/\beta$. Using this variable, the kinematic boundary of the reaction can be obtained by taking a limit  $D^2 \to 0$. For physical processes, the inequality $D^{2}\geq0$ must be satisfied.

To calculate convolutions of the form $[\epsilon^{\rho\alpha\beta\gamma}a_{\alpha}b_{\beta}c_{\gamma}h_{\rho}]^{2}$ in four dimensions or $[\epsilon^{\alpha\beta\gamma}a_{\alpha}b_{\beta}c_{\gamma}]^{2}$ in three dimensions, it is convenient to use the following identities
expressing the product of the Levi-Civita symbols through a determinant whose elements are the Kronecker delta symbols:

\begin{equation}
\epsilon^{\alpha\beta\gamma\rho}\epsilon_{\alpha_{1}\beta_{1}\gamma_{1}\rho_{1}} = -\left|
\begin{array}{cccc}
\delta^{\alpha}_{\alpha_{1}} & \delta^{\alpha}_{\beta_{1}} & \delta^{\alpha}_{\gamma_{1}} & \delta^{\alpha}_{\rho_{1}}\\
\delta^{\beta}_{\alpha_{1}} & \delta^{\beta}_{\beta_{1}} & \delta^{\beta}_{\gamma_{1}} & \delta^{\beta}_{\rho_{1}}\\
\delta^{\gamma}_{\alpha_{1}} & \delta^{\gamma}_{\beta_{1}} & \delta^{\gamma}_{\gamma_{1}} & \delta^{\gamma}_{\rho_{1}}\\
\delta^{\rho}_{\alpha_{1}} & \delta^{\rho}_{\beta_{1}} & \delta^{\rho}_{\gamma_{1}} & \delta^{\rho}_{\rho_{1}}
  \label{f:dim4}
\end{array}
\right|
\end{equation}

\begin{equation}
\epsilon^{\alpha\beta\gamma}\epsilon_{\alpha_{1}\beta_{1}\gamma_{1}} = \left|
\begin{array}{ccc}
\delta^{\alpha}_{\alpha_{1}} & \delta^{\alpha}_{\beta_{1}} & \delta^{\alpha}_{\gamma_{1}} \\
\delta^{\beta}_{\alpha_{1}} & \delta^{\beta}_{\beta_{1}} & \delta^{\beta}_{\gamma_{1}} \\
\delta^{\gamma}_{\alpha_{1}} & \delta^{\gamma}_{\beta_{1}} & \delta^{\gamma}_{\gamma_{1}} \\
\end{array}
\right|
  \label{f:dim3}
\end{equation}

These mathematical objects are well known in distance geometry \cite{a1}.
The proposed invariants for obtaining kinematic limits
are analogs of the Cayley-Menger determinant \cite{1}. This determinant
is calculated from a set of points (based on the distances between points). For any set of points in Euclidean space, it has a fixed sign (depending on the dimension of the determinant) or is equal to 0. If a set of points
produces a determinant of the wrong sign,
then they cannot belong to Euclidean space.
In our case, this corresponds to that the law of conservation of energy-momentum is not being satisfied, or the kinematic hypothesis
regarding the event is being invalid.
This determinant (with a known multiplicative factor) defines the square of the oriented volume of the simplex constructed from the vectors used to calculate the sign-definite invariant. Similarly, the square of the volume is expressed as
$[\epsilon^{\rho\alpha\beta\gamma}a_{\alpha}b_{\beta}c_{\gamma}h_{\rho}]^{2}$ or $[\epsilon^{\alpha\beta\gamma}a_{\alpha}b_{\beta}c_{\gamma}]^{2}$, depending on the number of vectors. The sign of the determinant (\ref{f:dim3}) calculated through the scalar products of four-vectors must be negative if the subspace based on these three four-vectors has the signature $(-;-;-)$. That is, it is a Euclidean space. Obviously, points in this space cannot belong to three-dimensional space-time. Therefore, the determinant becomes negative.

Checking an event for compatibility with the laws of conservation of energy-momentum
makes it necessary to calculate at least one invariant. To do this,
it must be parameterized based on the physical content of the process.
In principle, there can be several of these variables, in which case
all of them must be non-negative. In this case,
using coordinate systems (\ref{f:4}),(\ref{f:3}),
we can explicitly write down the energies and momenta of all
particles participating in the reaction and find the coordinates of the vectors in the corresponding bases.
This is necessary and sufficient for the law of conservation of energy and momentum to be satisfied.

Relation (\ref{f:metric10})
is also true in the four-dimensional case,
if $\vec{C}_{i}$ is a two-dimensional vector. If the subspace with a basis based on vectors a and b has signature $(+,-)$, then
  $\vec{C}^{2}$ is non-negative.  The determinant of the metric tensor for the above signature will be negative. This condition is expressed in the following form:

\begin{equation}
 \left|
\begin{array}{cc}
\frac{\displaystyle b^{2} } {\displaystyle a^{2}b^{2}-(ab)^{2} } & \frac{\displaystyle -ab } {\displaystyle a^{2}b^{2}-(ab)^{2} } \\
\frac{\displaystyle -ab } {\displaystyle a^{2}b^{2}-(ab)^{2} } & \frac{\displaystyle a^{2} } {\displaystyle a^{2}b^{2}-(ab)^{2} } \\
\end{array}
\right|=\frac{\displaystyle 1 } {\displaystyle a^{2}b^{2}-(ab)^{2} }<0.
\label{eq:met2}
\end{equation}

In other cases, if $a^{2}b^{2}-(ab)^{2}>0$, the sign of $\vec{C}^{2}$ is not fixed. If the determinant of the matrix
is negative, this means that the eigenvalues have different signs,
and the signature in the space of four-vectors $a$ and $b$ is $(+;-)$,
 while the signature
 in the space $\vec{C}$ is $(-;-)$.
 If the determinant is positive, the eigenvalues have the same sign,
and the signature in the space of four-vectors $a$ and $b$ is $(-;-)$, while the signature
in the space $\vec{C}$ is $(+;-)$ and the sign of $\vec{C}^{2}$ is not fixed.
The transition $\vec{C}^{2}\rightarrow0$ does not yield a kinematic limit.
To use expression (\ref{f:metric10}) to obtain a kinematic limit,
the condition $a^{2}b^{2}-(ab)^{2}<0$ must be satisfied.

\section{Examples of Application of the Kinematic Limits Method}

\begin{figure}[!h]
\centering
\includegraphics[width=.65\textwidth]{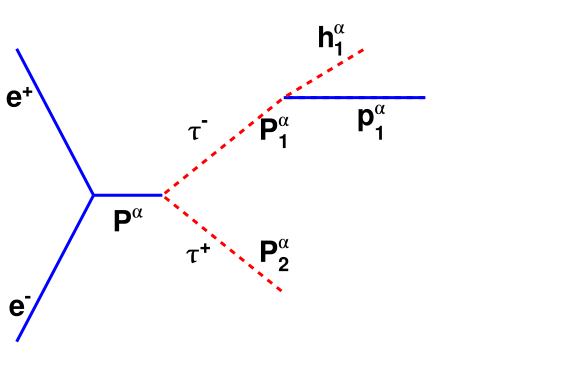}
\caption{\label{fig:M1M2p1} Schematic of the production and decay of tau leptons.
Solid lines indicate detected particles, dotted lines indicate undetected ones.}
\end{figure}

\subsection{Pseudomass Method for Measuring the $\tau$ Lepton Mass }
\label{sec:pm}

The pseudomass method was proposed by the ARGUS collaboration \cite{c} to measure the $\tau$ lepton mass. The dataset was obtained at the $e^{+}e^{-}$ collider at energies from 9.4 to 10.6 GeV.
Leptons were produced in pairs, one of which (for example, $\tau^{+}$) decayed into $e^{+}\bar{\nu}_{\tau}\nu_{e}$, the other into $\pi^{-}\pi^{-}\pi^{+}\nu_{\tau}$. The pseudomass $M_{P}$ is given by the following expression:

\begin{eqnarray}
M_{P}^{2}=m^{2}_{h}+2(E_{h}-\sqrt{E_{h}^{2}-m^{2}_{h}})(E_{b}-E_{h}),
\label{f:Argus}
\end{eqnarray}

Here $E_{b}$ is the beam energy, $E_{h}$ is the hadron ($\pi^{-}\pi^{-}\pi^{+}$) energy in the center-of-mass system, and $m_{h}$ is the hadron mass. We will now derive a more general formula and show how it can be reduced to the method proposed by the ARGUS collaboration.

The four-momentum $P$ of the initial state $e^{+}e^{-}$ is assumed to be known. This state then transitions to $\tau^{+}\tau^{-}$ with masses $M_{1}$ and $M_{2}$, which for generality we will assume to be different. These masses are unknown parameters that need to be measured.
The first $\tau$ lepton decays into two subsystems: a hadron state and a neutrino.

Hadrons are reconstructed, and their total four-momentum is $p_{1}$ ($p_{1}^{2}=m_{1}^{2}$), while the neutrino is not reconstructed and has a four-momentum of $h_{1}$ ($h_{1}^{2}=\mu^{2}_{1}$) (see Fig. (\ref{fig:M1M2p1})).
Let us find constraints on $M_{1}^{2}$ and $M_{2}^{2}$ for given $P$, $p_{1}$, and $\mu^{2}_{1}$. The scalar products $(Ph_{1})$, $(p_{1}h_{1})$
can be found as functions of $M^{2}_{1}$ and $M^{2}_{2}$. From the decay kinematics shown in Fig. (\ref{fig:M1M2p1}) it follows:

\begin{eqnarray}
  M_{1}^{2}=(p_{1}+h_{1})^{2}=m^{2}_{1}+2(p_{1}h_{1})+\mu^{2}_{1},\nonumber \\
    M_{2}^{2}=(P-p_{1}-h_{1})^2=P^{2}-2(Pp_{1})-2(Ph_{1})+M_{1}^{2}.
     \label{f:hp}
\end{eqnarray}

  A system of linear equations allows us to express the scalar products $(p_{1}h_{1})$ and
$(Ph_{1})$ in terms of the masses:

\begin{eqnarray}
(p_{1}h_{1})=\frac{\displaystyle M_{1}^{2}-m^{2}_{1}-\mu^{2}_{1}}{\displaystyle 2},\nonumber \\
(Ph_{1})=\frac{\displaystyle M_{1}^{2}-M^{2}_{2}+P^{2}-2(Pp_{1})}{\displaystyle 2}.
\label{f:hp1}
\end{eqnarray}

It is clear that the scalar products are linear functions of $M_{1}^{2}$ and $M_{2}^{2}$.
Now we can find $\mu^{2}_{1}$ as a quadratic form of $M_{1}^{2}$, $M_{2}^{2}$, and one free parameter $C_{1}=h_{1}^{\alpha}\epsilon_{\alpha\beta\gamma}P^{\beta}p_{1}^{\gamma}$.
To obtain a quadratic form in the principal axes, in the expression (\ref{f:metric10})
we need to take $u_{1}=u_{2}=h_{1}$, $a^{\alpha}=P^{\alpha}$, $b^{\alpha}=\frac{(Pp_{1})}{P^{2}}P^{\alpha}-p_{1}^{\alpha}$ and
$C_{2}=C_{1}$. From (Eq.\ref{f:hp1}), we obtain:

\begin{eqnarray}
\frac{\displaystyle (Pp_{1})}{\displaystyle P^{2}}(Ph_{1})-(p_{1}h_{1})=\frac{\displaystyle (Pp_{1})}{\displaystyle P^{2}}\frac{\displaystyle M_{1}^{2}-M^{2}_{2}+P^{2}-2(Pp_{1})}{\displaystyle 2}-\frac{\displaystyle M_{1}^{2}-m^{2}_{1}-\mu^{2}_{1}}{\displaystyle 2}.
\label{f:hpp}
\end{eqnarray}

Substituting  Eqs.(\ref{f:hp1}) and (\ref{f:hpp}) to Eq.(\ref{f:metric10}), we obtain the quadratic equation as a function of $M_{1}^2$ and $M_{2}^{2}$.

\begin{eqnarray}
\frac{\displaystyle [\frac{(Pp_{1})}{P^{2}}(Ph_{1})-(p_{1}h_{1})]^{2}}{\displaystyle [-(Pp_{1})^{2}/P^{2}+m^{2}_{1}]}
+\frac{\displaystyle (Ph_{1})^{2} }{\displaystyle P^{2}}+\frac{\displaystyle [h_{1}^{\alpha}\epsilon_{\alpha\beta\gamma}P^{\beta}p_{1}^{\gamma}]^{2}}{\displaystyle -(Pp_{1})^{2}+m^{2}_{1}P^{2}}=\nonumber\\ 
=\frac{\displaystyle [(Pp_{1})[M_{1}^{2}-M^{2}_{2}+P^{2}-2(Pp_{1})]/P^{2}-M_{1}^{2}+m^{2}_{1}+\mu^{2}_{1}]^{2}}{\displaystyle 4[-(Pp_{1})^{2}/P^{2}+m^{2}_{1}]}+\nonumber\\ 
+\frac{\displaystyle [M_{1}^{2}-M^{2}_{2}+P^{2}-2(Pp_{1})]^{2} }{\displaystyle 4P^{2}}+\frac{\displaystyle C^{2}_{1}}{\displaystyle -(Pp_{1})^{2}+m^{2}_{1}P^{2}}=\mu^{2}_{1}. 
\label{f:hpz}
\end{eqnarray}

This is the equation of a hyperbola in the plane $(M^{2}_{1};M^{2}_{2})$, since
the factors of the perfect squares $[\frac{(Pp_{1})}{P^{2}}(Ph_{1})-(p_{1}h_{1})]^{2}$
and $[h_{1}^{\alpha}\epsilon_{\alpha\beta\gamma}P^{\beta}p_{1}^{\gamma}]^{2}$
are negative, and the factor of the perfect square $(Ph_{1})^{2}$ is positive.

Substituting into (\ref{f:hpz}) the neutrino mass $\mu^{2}_{1}=0$:

\begin{eqnarray}
\frac{\displaystyle [M_{1}^{2}-M^{2}_{2}+P^{2}-2(Pp_{1})]^{2} }{\displaystyle 4P^{2}}+\nonumber\\
+\frac{\displaystyle [(Pp_{1})[M_{1}^{2}-M^{2}_{2}+P^{2}-2(Pp_{1})]/P^{2}-M_{1}^{2}+m^{2}_{1}]^{2}}{\displaystyle 4[-(Pp_{1})^{2}/P^{2}+m^{2}_{1}]}+\nonumber\\
+\frac{\displaystyle C^{2}_{1}}{\displaystyle -(Pp_{1})^{2}+m^{2}_{1}P^{2}}=0.
\label{f:cone}
\end{eqnarray}

Since the neutrino mass is zero at the energy scale under consideration, its equations of motion
represent a light cone.
From equation (\ref{f:cone})
$M^{2}_{1}$ and $M^{2}_{2}$ can be found as functions of the free parameters $\lambda$ and $\Theta$, which define
the cone equations parametrically ($\lambda\in(0,\infty)$, $\Theta\in[0,\pi]$):

\begin{eqnarray}
  M_{1}^{2}-M^{2}_{2}+P^{2}-2Pp_{1}=2\sqrt{P^{2}}\lambda
\label{f:M1M2lam0}
\end{eqnarray}

\begin{eqnarray}
Pp_{1}[M_{1}^{2}-M^{2}_{2}+P^{2}-2Pp_{1}]/P^{2}-M_{1}^{2}+m^{2}_{1}= 2\sqrt{Pp_{1}^{2}/P^{2}-m^{2}_{1}}\cos{\Theta}\lambda
\label{f:M1M2lam1}
\end{eqnarray}

\begin{eqnarray}
\frac{\displaystyle C^{2}_{1}}{\displaystyle -(Pp_{1})^{2}+m^{2}_{1}P^{2}}=-\lambda^{2}\sin^{2}{\Theta} 
\label{f:M1M2lam2}
\end{eqnarray}

Solving the linear equation
gives $M^{2}_{1}$ and $M^{2}_{2}$:

\begin{eqnarray}
M_{1}^{2}=m^{2}_{1}+2[(Pp_{1})/\sqrt{P^{2}}-\sqrt{(Pp_{1})^{2}/P^{2}-m^{2}_{1}}\cos{\Theta}]\lambda,\nonumber\\
M_{2}^{2}=m^{2}_{1}+P^{2}-2Pp_{1}+2\Big[(Pp_{1})/\sqrt{P^{2}}-\sqrt{P^{2}}-\sqrt{(Pp_{1})^{2}/P^{2}-m^{2}_{1}}\cos{\Theta}\Big]\lambda.
\label{f:sM1M2lam}
\end{eqnarray}

\begin{figure}[!h]
\centering
\includegraphics[width=.65\textwidth]{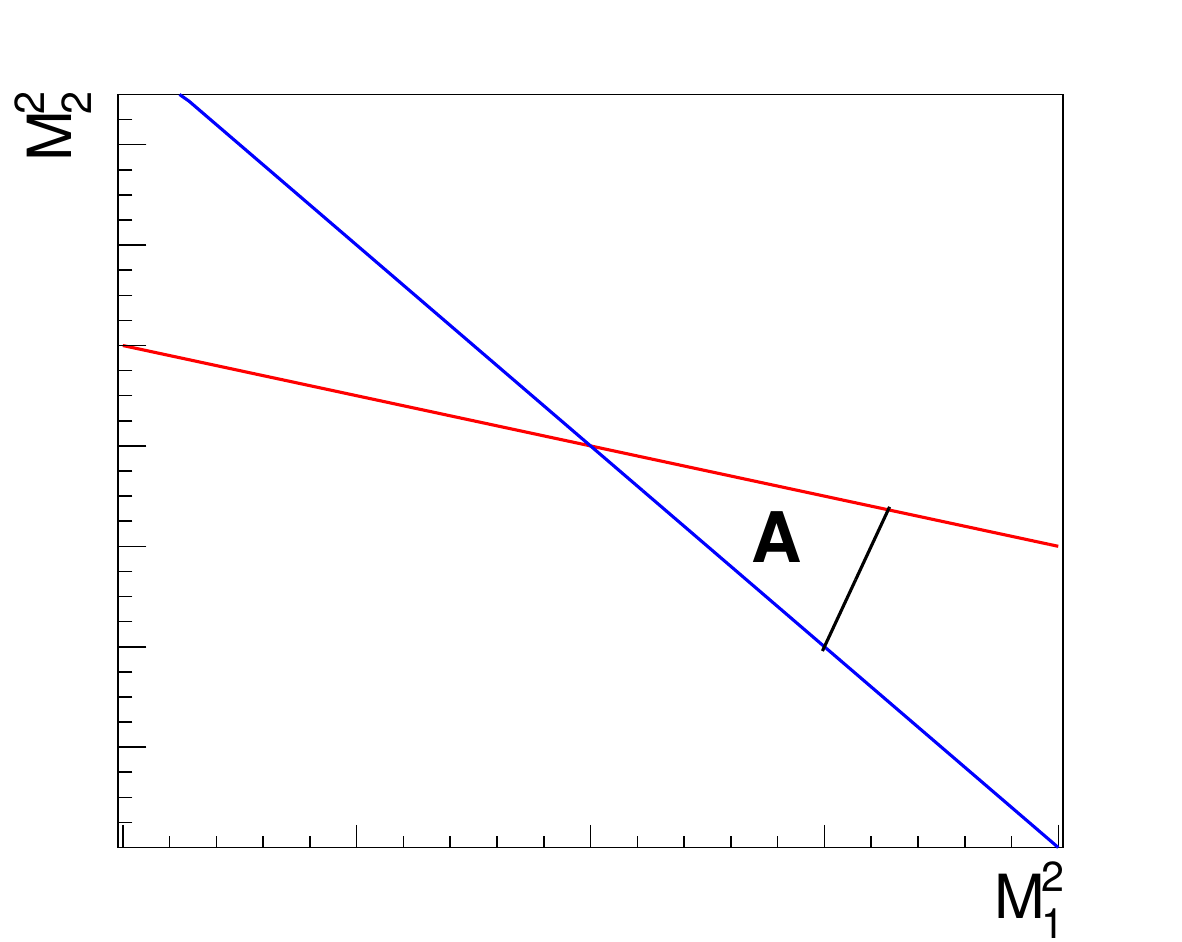}
\caption{\label{fig:sector} Kinematic constraints on $M_{1}^{2}$ and $M_{2}^{2}$. The blue line corresponds to $\cos{\Theta}=1$, the red line corresponds to $\cos{\Theta}=-1$. Triangle "A" is the physically accessible region. The crossing point of blue and red lines corresponds to $\lambda=0$.}
\end{figure}

By changing the values of the parameters $\lambda$ and $\Theta$,
we can reach any point inside the two-dimensional light cone in space ($M^{2}_{1};M^{2}_{2}$) Fig. (\ref{fig:sector}).
Transition to the kinematic limit implies $C_{1}=0$ or $\cos{\Theta}=\pm1$. The different signs define two directions of neutrino motion in a two-dimensional space based on the four-vectors
$P$, $p_{1}$.
The parameter $\lambda$ is the neutrino energy in the $e^{+}e^{-}$ center-of-mass system.
As $\lambda$ increases, the scalar product $(Pp_{1})+(Ph_{1})$ becomes equal to $P^{2}$, and the scalar product $(PP_{2})$ is zero.
As $\lambda$ increases further, $(PP_{2})$ becomes negative. This corresponds to the process where
$\tau^{+}$ is transferred from the final to the initial state. 
The result is a triangular region "A" as a constraint on the parameters $M^{2}_{1}$ and $M^{2}_{2}$.

Using the pseudomass method, the OPAL, Belle, and BABAR collaborations obtained a constraint on the mass difference between $\tau^{+}$ and
$\tau^{-}$ \cite{b}.
In these studies, the pseudomass $\tau^{+}$ was obtained under the assumption
that the masses of $\tau^{+}$ and
$\tau^{-}$ were equal. The mass of $\tau^{-}$ was measured similarly. The result is then presented as a measurement of the mass difference between $\tau^{+}$ and $\tau^{-}$. But from equation (\ref{f:sM1M2lam}) it is clear that only a certain region, namely triangle "A" in the plane
$(M_{1}^{2};M_{2}^{2})$ or $(M_{\tau^{-}}^{2};M_{\tau^{+}}^{2})$, can be localized.
The situation differs fundamentally from the case where the two masses are equal
$M_{\tau^{-}}^{2}=M_{\tau^{+}}^{2}$. 

To obtain the pseudomass (\ref{f:Argus}) from \cite{c} in more familiar notation,
substitute $P^{2}=4E_{b}^{2}$, $(Pp_{1})=2E_{b}E_{h}$,
$\cos{\Theta}=1,m_{1}=m_{h}$, and $\lambda=E_{b}-E_{h}$ into (\ref{f:sM1M2lam})
(these values are obtained from the condition $M_{p}^{2}=M^{2}_{1}=M^{2}_{2}$).

\subsection{Pseudomass Method for Measuring the Mass of the $W$ Boson}

\begin{figure}[!h]
\centering
\includegraphics[width=.65\textwidth]{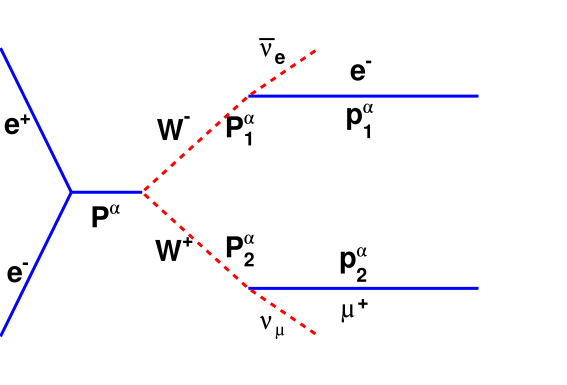}
\caption{\label{fig:PM} Schematic diagram of the production and decay of a pair of $W^{+}W^{-}$ bosons used in the pseudomass method.
Solid lines indicate detected particles, dotted lines indicate the undetected ones.}
\end{figure}

In paper \cite{100}, the authors introduced a method for measuring the mass of $W^{\pm}$
in the process $e^{+}e^{-}\to W^{+}W^{-}\to e^{-}+\bar{\nu}_{e}+\mu^{+}+\nu_{\mu}$.
It is proposed to measure the mass by registering $e^{-}$ and $\mu^{+}$,
studying the correlations between the emission angles of the registered particles.
Let us consider this problem within the framework of our general approach. We denote
$p_{1}=p_{e^{-}}$,
$p_{2}=p_{\mu^{+}}$,
$P_{1}=p_{W^{-}}$, $P_{2}=p_{W^{+}}$, and $P=p_{W^{-}}+p_{W^{+}}$.
For generality, we leave free the squares of the neutrino mass - $\mu_{1}^{2}$
and $\mu_{2}^{2}$ for $\bar{\nu}_{e}$ and $\nu_{\mu}$, and the masses $M^{2}_{1}$, $M^{2}_{2}$ for $W^{+}$ and $W^{-}$. At the same time,

\begin{eqnarray}
P_{1}^{2}=M_{1}^{2},\nonumber \\
P_{2}^{2}=M_{2}^{2}, \nonumber \\
P^{\alpha}=P_{1}^{\alpha}+P_{2}^{\alpha}.
\label{f:Pcon}
\end{eqnarray}

Now we can express $P_{1}$ and $P_{2}$ in terms of $P$ and a new four-vector $Q$
orthogonal to $P$ as:

\begin{eqnarray}
P_{1}^{\alpha}=\frac{\displaystyle P^{2}+M^{2}_{1}-M^{2}_{2}}{\displaystyle 2P^{2}}P^{\alpha}+\frac{\displaystyle Q^{\alpha}}{2},\nonumber \\
P_{2}^{\alpha}=\frac{\displaystyle P^{2}+M^{2}_{2}-M^{2}_{1}}{\displaystyle 2P^{2}}P^{\alpha}-\frac{\displaystyle Q^{\alpha}}{2}. 
\label{f:PQ}
\end{eqnarray}

In this case:

\begin{eqnarray} 
(PQ)=0,\nonumber \\ 
Q^{2}=2M^{2}_{1}+2M^{2}_{2}-P^{2}-\frac{\displaystyle (M_{1}^{2}-M^{2}_{2})^{2}}{P^{2}}. 
\label{f:QCon}
\end{eqnarray}

The scalar products $(Qp_{1})$ and $(Qp_{2})$
can be found from the equations:

\begin{eqnarray}
(P_{1}-p_{1})^{2}=\mu^{2}_{1}
=M^{2}_{1}-\frac{\displaystyle P^{2}+M^{2}_{1}-M^{2}_{2}}{\displaystyle P^{2}}(Pp_{1})-(Qp_{1})+m_{1}^{2},\nonumber \\
(P_{2}-p_{2})^{2}=\mu^{2}_{2}
=M^{2}_{2}-\frac{\displaystyle P^{2}+M^{2}_{2}-M^{2}_{1}}{\displaystyle P^{2}}(Pp_{2})+(Qp_{2})+m_{2}^{2}. 
\label{f:Qp1}
\end{eqnarray}

Solving the system of linear equations, we obtain:

\begin{eqnarray}
(Qp_{1})=M^{2}_{1}-\frac{\displaystyle P^{2}+M^{2}_{1}-M^{2}_{2}}{\displaystyle P^{2}}(Pp_{1})+m_{1}^{2}-\mu^{2}_{1}, \nonumber \\
(Qp_{2})=-M^{2}_{2}+\frac{\displaystyle P^{2}+M^{2}_{2}-M^{2}_{1}}{\displaystyle P^{2}}(Pp_{2})-m_{2}^{2}+\mu^{2}_{2}.
\label{f:Qp2}
\end{eqnarray}

To find $Q^{2}$, we make a substitution in (\ref{f:metric20}):
$(v_{1}v_{2})=Q^{2}$, $a=P$, $b=p_{1}$,
$c=p_{2}$, $A_{1}=A_{2}=(QP)=0$,
$B_{1}=B_{2}=(Qp_{1})$ from (\ref{f:Qp2}),
$C_{1}=C_{2}=(Qp_{2})$ from (\ref{f:Qp2}),
$D=Q_{\rho}\epsilon^{\rho\alpha\beta\gamma}P_{\alpha}p_{1\beta}p_{2\gamma}$.
These substitutions give the following equation:

\begin{eqnarray}
  D^{2}+(-P^{2}g_{\nu\sigma}+P_{\nu}P_{\sigma})\times \nonumber \\ 
  \times[(M^{2}_{1}-\frac{\displaystyle P^{2}+M^{2}_{1}-M^{2}_{2}}{\displaystyle P^{2}}(Pp_{1})+m^{2}_{1}-\mu^{2}_{1})p_{2}^{\nu}+(M^{2}_{2}-\frac{\displaystyle P^{2}+M^{2}_{2}-M^{2}_{1}}{\displaystyle P^{2}}(Pp_{2})+m^{2}_{2}-\mu^{2}_{2})p_{1}^{\nu}]\times \nonumber \\
\times[(M^{2}_{1}-\frac{\displaystyle P^{2}+M^{2}_{1}-M^{2}_{2}}{\displaystyle P^{2}}(Pp_{1})+m^{2}_{1}-\mu^{2}_{1})p_{2}^{\sigma}+(M^{2}_{2}-\frac{\displaystyle P^{2}+M^{2}_{2}-M^{2}_{1}}{\displaystyle P^{2}}(Pp_{2})+m^{2}_{2}-\mu^{2}_{2})p_{1}^{\sigma}]=\nonumber \\
=[2M^{2}_{1}+2M^{2}_{2}-P^{2}-\frac{\displaystyle (M_{1}^{2}-M^{2}_{2})^{2}}{P^{2}}]\epsilon^{\alpha\beta\gamma\rho}\epsilon_{\alpha_{1}\beta_{1}\gamma_{1}\rho}P_{\alpha}p_{1\beta}p_{2\gamma}P^{\alpha_{1}}p_{1}^{\beta_{1}}p_{2}^{\gamma_{1}}.\nonumber \\
\label{eq:main2}
\end{eqnarray}

The kinematic limit is reached at $D^{2}=0$.
Let us show that equation (\ref{eq:main2}) is a quadratic form in the variables $M_{1}^{2},M_{2}^{2}$, corresponding to an ellipse.
We will do this indirectly by obtaining the same equations in a different way.
To do this, we need to split the $P-p_{1}-p_{2}$ system into two neutrinos with squared invariant masses
$\mu_{1}^{2}$ and $\mu_{2}^{2}$. And rewrite the relations (\ref{f:PQ}), (\ref{f:QCon})
replacing $P$ with $P-p_{1}-p_{2}$,
$M_{1}^{2}$ with $\mu^{2}_{1}$, $M_{2}^{2}$ with $\mu^{2}_{2}$, $P_{1}$ with $h_{1}$,
$P_{2}$ with $h_{1}$, and $Q$ with $q$:

\begin{eqnarray}
h_{1}^{\alpha}=\frac{\displaystyle (P-p_{1}-p_{2})^{2}+\mu^{2}_{1}-\mu^{2}_{2}}{\displaystyle 2(P-p_{1}-p_{2})^{2}}[P-p_{1}-p_{2}]^{\alpha}+\frac{\displaystyle q^{\alpha}}{2},\nonumber \\
h_{2}^{\alpha}=\frac{\displaystyle (P-p_{1}-p_{2})^{2}+\mu^{2}_{2}-\mu^{2}_{1}}{\displaystyle 2(P-p_{1}-p_{2})^{2}}[P-p_{1}-p_{2}]^{\alpha}-\frac{\displaystyle q^{\alpha}}{2}, \nonumber \\ 
\label{f:h}
\end{eqnarray}

Where

\begin{eqnarray}
(Pq)-(p_{1}q)-(p_{2}q)=0,\nonumber \\ 
q^{2}=2\mu^{2}_{1}+2\mu^{2}_{2}-(P-p_{1}-p_{2})^{2}-\frac{\displaystyle (\mu_{1}^{2}-\mu^{2}_{2})^{2}}{(P-p_{1}-p_{2})^{2}}. 
\label{f:Qnew}
\end{eqnarray}

Now we can find $q^{2}$ as a function of $(qp_{1})$, $(qp_{2})$, $(qP)-(qp_{1})-(qp_{2})$
and $D^{\prime}=\epsilon^{\alpha\beta\gamma\rho}q_{\alpha}[P-p_{1}-p_{2}]_{\beta}p_{1\gamma}p_{2\rho}$.
The last parameter is free, and $(qP)-(qp_{1})-(qp_{2})$ is zero. Here we assume that $a=P-p_{1}-p_{2}$,
$b=p_{1}$, $c=p_{2}$ in (\ref{f:metric20}).

\begin{eqnarray} 
D^{\prime 2}+[-(P-p_{1}-p_{2})^{2}g_{\nu\sigma}+[P-p_{1}-p_{2}]_{\nu}[P-p_{1}-p_{2}]_{\sigma}]\times \nonumber \\ 
\times[(qp_{1})p_{2}^{\nu}-(qp_{2})p_{1}^{\nu}][(qp_{1})p_{2}^{\sigma}-(qp_{2})p_{1}^{\sigma}]=\nonumber \\ 
=[2\mu^{2}_{1}+2\mu^{2}_{2}-(P-p_{1}-p_{2})^{2}-\frac{\displaystyle (\mu_{1}^{2}-\mu^{2}_{2})^{2}}{\displaystyle (P-p_{1}-p_{2})^{2}}]\times \nonumber \\
\times\epsilon^{\alpha\beta\gamma\rho}\epsilon_{\alpha_{1}\beta_{1}\gamma_{1}\rho}[P-p_{1}-p_{2}]_{\alpha}p_{1\beta}p_{2\gamma}[P-p_{1}-p_{2}]^{\alpha_{1}}p_{1}^{\beta_{1}}p_{2}^{\gamma_{1}}.
\label{eq:mainz}
\end{eqnarray}

Note that $D^{\prime}=D$, and equation (\ref{eq:mainz}) is a rewritten equation (\ref{eq:main2}). Instead of $M^{2}_{1}$ and $M^{2}_{2}$, the variables $(qp_{1})$ and $(qp_{2})$ are used.
This is the equation of ellipse for the variables $(qp_{1})$ and $(qp_{2})$. Since $q^{2}<0$, $q^{2}$ as a function of the coordinates ($(qp_{1})$,$(qp_{2})$,$\epsilon^{\alpha\beta\gamma\rho}q_{\alpha}[P-p_{1}-p_{2}]_{\beta}p_{1\gamma}p_{2\rho}$) of the four-vector $q$
represents a quadratic form with an inertia index -3. The maximum size of the ellipse is achieved when $D^{\prime}=0$.
We express $M_{1}^{2}$ and $M_{2}^{2}$ as functions of $(qp_{1})$, $(qp_{2})$, and
the reconstructed four-vectors:

\begin{eqnarray}
M^{2}_{1}=(h_{1}+p_{1})^{2}=(\frac{\displaystyle (P-p_{1}-p_{2})^{2}+\mu^{2}_{1}-\mu^{2}_{2}}{\displaystyle 2(P-p_{1}-p_{2})^{2}}[P-p_{1}-p_{2}]^{\alpha}+\frac{\displaystyle q^{\alpha}}{2}+p_{1}^{\alpha})^{2}=\nonumber \\ 
=\mu^{2}_{1}+\frac{\displaystyle (P-p_{1}-p_{2})^{2}+\mu^{2}_{1}-\mu^{2}_{2}}{\displaystyle 2(P-p_{1}-p_{2})^{2}}[(Pp_{1})-m_{1}^{2}-(p_{2}p_{1})]+qp_{1}+m_{1}^{2},\nonumber \\ 
M^{2}_{2}=(h_{2}+p_{2})^{2}=(\frac{\displaystyle (P-p_{1}-p_{2})^{2}+\mu^{2}_{2}-\mu^{2}_{1}}{\displaystyle 2(P-p_{1}-p_{2})^{2}}[P-p_{1}-p_{2}]^{\alpha}-\frac{\displaystyle q^{\alpha}}{2}+p_{2}^{\alpha})^{2}=\nonumber \\ 
=\mu^{2}_{2}+\frac{\displaystyle (P-p_{1}-p_{2})^{2}+\mu^{2}_{2}-\mu^{2}_{1}}{\displaystyle 2(P-p_{1}-p_{2})^{2}}[(Pp_{2})-(p_{2}p_{1})-m_{2}^{2}]-qp_{2}+m_{2}^{2}.\nonumber \\
\label{eq:M}
\end{eqnarray}

Thus, $M_{1}^{2}$ is a linear function of $(qp_{1})$, and $M_{2}^{2}$ is a linear function of $(qp_{2})$.
All other parameters in the equation (\ref{eq:M}) are known.

To obtain the pseudomass in the case of $e^{+}e^{-}\to W^{+}W^{-}\to e^{-}+\bar{\nu}_{e}+\mu^{+}+\nu_{\mu}$, we need to substitute $M_{1}=M_{2}=M_{W}$,
and $\mu^{2}_{1}=\mu^{2}_{2}=0$ Eq.(\ref{eq:mainz}) reads.

\begin{eqnarray} 
(-P^{2}g_{\nu\sigma}+P_{\nu}P_{\sigma})\times \nonumber \\
\times[[M^{2}_{W}+m_{1}^{2}-(Pp_{1})]p_{2}^{\nu}+[M^{2}_{W}+m_{2}^{2}-(Pp_{2})]p_{1}^{\nu}]\times \nonumber \\
\times[[M^{2}_{W}+m_{1}^{2}-(Pp_{1})]p_{2}^{\sigma}+[M^{2}_{W}+m_{2}^{2}-(Pp_{2})]p_{1}^{\sigma}]= \nonumber \\
=[4M^{2}_{W}-P^{2}]\epsilon^{\alpha\beta\gamma\rho}\epsilon_{\alpha_{1}\beta_{1}\gamma_{1}\rho}P_{\alpha}p_{1\beta}p_{2\gamma}P^{\alpha_{1}}p_{1}^{\beta_{1}}p_{2}^{\gamma_{1}}.
\label{eq:main2p}
\end{eqnarray}

The substitution yields a quadratic equation (\ref{eq:main2p}) on $M^{2}_{W}$,
from which two pseudomass values are obtained. The square of the $W$ boson mass must be between the two roots in a form $M^{2}(root_{1})\leq M^{2}_{W}\leq M^{2}(root_{2})$. The authors in \cite{100} proposed studying correlations
between the angles $\mu^{+}$ and $e^{-}$ to measure the $W^{+}$ mass, assuming that the $W^{+}$ and $W^{-}$ masses are equal. Solving the
quadratic equation (\ref{eq:main2p}) yields two pseudomasses.
This is direct way to measure $W$ boson mass.

\subsection{ Kinematic limits for searches for the second-class currents in the $\tau^{-}\to\pi^{-}\eta\nu_{\tau}$ decay }

The upper limit on the $\tau^{-}\to\pi^{-}\eta\nu_{\tau}$ decay probability is $9.9\times10^{-5}$ \cite{secclascur} and was obtained using the statistics of $4.32\times10^{8}$ pairs of $\tau^{+}\tau^{-}$ leptons. It can be seen that it is significantly
bigger than the reciprocal of the total number of $\tau$ leptons. This is due to the high background level. If the number of background events is much greater than one,
then the sensitivity is proportional to $1/\sqrt{N_{\tau\tau}}$, where $N_{\tau\tau}$ is the number of $\tau^{+}\tau^{-}$ pairs. When the number of background events is much less
than one, the sensitivity is proportional to $1/N_{\tau\tau}$. In work \cite{secclascur}, the number of background events
numbers is in the hundreds. One search strategy
is to identify a region containing signal events but no
background events. We will show how to find a region in which the signal and background
do not intersect.

The main background for studying $\tau^{-}\to\pi^{-}\eta\nu_{\tau}$ comes from the decays
$\tau^{-}\to\rho^{-}\eta\nu_{\tau}\to\pi^{-}\pi^{0}\eta\nu_{\tau}$ with lost
$\pi^{0}$ and $\tau^{-}\to\pi^{-}K^{0}\eta\nu_{\tau}$ with lost $K^{0}$. We will find necessary and sufficient conditions for an event to satisfy the kinematics of the signal process and the two background types.

\subsubsection{ Conditions on the signal process }

We begin with equation (\ref{eq:main2}), in which we set $M_{1}=M_{2}=m_{\tau}$,
assuming that the four-momentum $P$ corresponds to the initial state, $p_{1}$ corresponds to the $\pi^{-}\eta$ system from
the $\tau^{-}$ decay, and $p_{2}$ corresponds to the detected $\tau^{+}$ decay products. The four-momentum $Q$, as in the previous case,
is equal to the difference between the four-momenta $\tau^{-}$ and $\tau^{+}$. $\mu^{2}_{1}$ is the invariant mass of $\nu_{\tau}$ from the decay of $\tau^{-}$, where $\mu^{2}_{2}$ is the invariant mass
of an antineutrino if the tagging $\tau^{+}$ decays semileptonically, or
of a neutrino-antineutrino pair if $\tau^{+}$ decays leptonically.
The parameters $\mu^{2}_{1}$
and $\mu^{2}_{2}$ remain free. As a result, the equation (\ref{eq:main2})
 takes the form:

\begin{eqnarray} 
D^{2}+(-P^{2}g_{\nu\sigma}+P_{\nu}P_{\sigma})\times \nonumber \\
\times[[m^{2}_{\tau}-\mu_{1}^{2}+m_{1}^{2}-(Pp_{1})]p_{2}^{\nu}+[m^{2}_{\tau}-\mu_{2}^{2}+m_{2}^{2}-(Pp_{2})]p_{1}^{\nu}]\times \nonumber \\
\times[[m^{2}_{\tau}-\mu_{1}^{2}+m_{1}^{2}-(Pp_{1})]p_{2}^{\sigma}+[m^{2}_{\tau}-\mu_{2}^{2}+m_{2}^{2}-(Pp_{2})]p_{1}^{\sigma}]= \nonumber \\
=[4m^{2}_{\tau}-P^{2}]\epsilon^{\alpha\beta\gamma\rho}\epsilon_{\alpha_{1}\beta_{1}\gamma_{1}\rho}P_{\alpha}p_{1\beta}p_{2\gamma}P^{\alpha_{1}}p_{1}^{\beta_{1}}p_{2}^{\gamma_{1}}.
\label{eq:main3}
\end{eqnarray}

For the signal process $\tau^{-}\to\pi^{-}\eta\nu_{\tau}$, necessary and sufficient conditions can be found from the fulfillment of the energy-momentum conservation law. In equation (\ref{eq:main3}), if
$\tau^{-}$ decays to $\pi^{-}\eta\nu_{\tau}$,
$D^{2}$ takes a non-negative value
at the point: $\mu^{2}_{1}=0$ and $\mu^{2}_{2}=0$.
Here, it is assumed that all decay products of $\tau$-leptons
except neutrinos and $\pi^{0}$ or $K^{0}$ in background processes
are reconstructed.
The inequality $D^{2}(\mu^{2}_{1}=\mu^{2}_{2}=0)\geq0$ has the following form:

\begin{eqnarray}
\label{eq:Fcond}
(m^{2}_{\tau}+m_{1}^{2}-(Pp_{1}))^{2}[P^{2}m_{2}^{2}-(Pp_{2})^{2}]+2(m^{2}_{\tau}+m_{1}^{2}-(Pp_{1}))(m^{2}_{\tau}+m_{2}^{2}-(Pp_{2}))\times\nonumber \\
\times[P^{2}(p_{1}p_{2})-(Pp_{1})(Pp_{2})]+(m^{2}_{\tau}+m_{2}^{2}-(Pp_{2}))^{2}[P^{2}m_{1}^{2}-(Pp_{1})^{2}]+\nonumber \\
+(4m^{2}_{\tau}-P^{2})\epsilon^{\alpha\beta\gamma\rho}\epsilon_{\alpha_{1}\beta_{1}\gamma_{1}\rho}P_{\alpha}p_{1\beta}p_{2\gamma}P^{\alpha_{1}}p_{1}^{\beta_{1}}p_{2}^{\gamma_{1}}\geq0.\nonumber \\
\end{eqnarray}

If inequality (\ref{eq:Fcond})
is not true, and $\tau^{+}$ that is to decays into leptons,
there is another way to satisfy the law of energy conservation, achieve a non-negative value of $D^{2}$ in some domain $\mu^{2}_{2}>0$, with $\mu^{2}_{1}=0$. We find an expression for the maximum value of roots $\mu^{2}_{2}(+)$ and minimum value of roots $\mu^{2}_{2}(-)$ of the quadratic equation see figure \ref{fig:ellipse} for the definition $\mu^{2}_{2}(+),\mu^{2}_{2}(-)$:

\begin{eqnarray}
\label{eq:q2}
D^{2}(\mu^{2}_{1}=0;\mu^{2}_{2})=0
\end{eqnarray}
Here $D^{2}$ is defined by equation (\ref{eq:main3}). Let's introduce new variables $x_{1}$ and $x_{2}$:

\begin{eqnarray}
  \label{eq:shift}
  \mu^{2}_{1}=m^{2}_{\tau}+m^{2}_{1}-(Pp_{1})+\sqrt{P^2-4m^{2}_{\tau}}\sqrt{(Pp_{1})^{2}/P^{2}-m_{1}^{2}}x_{1}, \nonumber \\
  \mu^{2}_{2}=m^{2}_{\tau}+m^{2}_{2}-(Pp_{2})+\sqrt{P^2-4m^{2}_{\tau}}\sqrt{(Pp_{2})^{2}/P^{2}-m_{2}^{2}}x_{2},
  \end{eqnarray}
and an independent invariant $\cos{\Theta}$, defined as:

\begin{eqnarray}
\label{eq:cos}
\cos{\Theta}=\frac{\displaystyle (Pp_{1})(Pp_{2})/P^{2}-(p_{1}p_{2}) }{\displaystyle \sqrt{(Pp_{1})^{2}/P^{2}-m_{1}^{2}}\sqrt{(Pp_{2})^{2}/P^{2}-m_{2}^{2}} }.
\end{eqnarray}

Now we can see that the right side of Eq.(\ref{eq:main3}).

\begin{eqnarray}
  \label{eq:eps}
  \epsilon^{\alpha\beta\gamma\rho}\epsilon_{\alpha_{1}\beta_{1}\gamma_{1}\rho}P_{\alpha}p_{1\beta}p_{2\gamma}P^{\alpha_{1}}p_{1}^{\beta_{1}}p_{2}^{\gamma_{1}}= \nonumber \\
 =-P^{2}[(Pp_{1})^{2}/P^{2}-m^{2}_{1}][(Pp_{2})^{2}/P^{2}-m^{2}_{2}]\sin^{2}{\Theta}.
  \end{eqnarray}
Indeed:

\begin{eqnarray} 
\label{eq:a} 
-P^{2}[(Pp_{1})^{2}/P^{2}-m^{2}_{1}][(Pp_{2})^{2}/P^{2}-m^{2}_{2}]\sin^{2}{\Theta}= \nonumber \\ 
=-P^{2}[(Pp_{1})^{2}/P^{2}-m^{2}_{1}][(Pp_{2})^{2}/P^{2}-m^{2}_{2}][1-\frac{\displaystyle [(Pp_{1})(Pp_{2})/P^{2}-(p_{1}p_{2})]^{2} }{\displaystyle [(Pp_{1})^{2}/P^{2}-m_{1}^{2}][(Pp_{2})^{2}/P^{2}-m_{2}^{2}] }]= \nonumber \\ 
=-[(Pp_{1})^{2}/P^{2}-m^{2}_{1}][(Pp_{2})^{2}-m^{2}_{2}P^{2}]+P^{2}[(Pp_{1})(Pp_{2})/P^{2}-(p_{1}p_{2})]^2= \nonumber \\ 
=-(Pp_{1})^{2}(Pp_{2})^{2}/P^{2}+m^{2}_{1}(Pp_{2})^{2}+m^{2}_{2}(Pp_{1})^{2}-P^{2}m^{2}_{1}m^{2}_{2}+ \nonumber \\ 
+(Pp_{1})^{2}(Pp_{2})^{2}/P^{2}-2(Pp_{1})(Pp_{2})(p_{1}p_{2})+P^{2}(p_{1}p_{2})^2=\nonumber \\
=m^{2}_{1}(Pp_{2})^{2}+m^{2}_{2}(Pp_{1})^{2}-P^{2}m^{2}_{1}m^{2}_{2}
-2(Pp_{1})(Pp_{2})(p_{1}p_{2})+P^{2}(p_{1}p_{2})^2.\nonumber \\
\end{eqnarray}
Then the left side of the equation (\ref{eq:main3}) can be rewritten as:

\begin{eqnarray}
\label{eq:mainlima} 
D^{2}+(P^2-4m^{2}_{\tau})(-P^{2}g_{\nu\sigma}+P_{\nu}P_{\sigma})[\sqrt{(Pp_{1})^{2}/P^{2}-m_ {1}^{2}}x_{1}p_{2}^{\nu}+\sqrt{(Pp_{2})^{2}/P^{2}-m_{2}^{2}}x_{2}p_{1}^{\nu}]\times\nonumber \\
\times[\sqrt{(Pp_{1})^{2}/P^{2}-m_{1}^{2}}x_{1}p_{2}^{\sigma}+\sqrt{(Pp_{2})^{2}/P^{2}-m_{2}^{2}}x_{2}p_{1}^{\sigma}]=\nonumber \\
=D^{2}+[P^2-4m^{2}_{\tau}]P^{2}[(Pp_{1})^{2}/P^{2}-m_{1}^{2}][(Pp_{2})^{2}/P^{2}-m_{2}^{2}]\times \nonumber\\
\times[x_{1}^{2}+2x_{1}x_{2}\frac{\displaystyle (Pp_{1})(Pp_{2})/P^{2}-(p_{1}p_{2}) }{\displaystyle \sqrt{(Pp_{1})^{2}/P^{2}-m_{1}^{2}}\sqrt{(Pp_{2})^{2}/P^{2}-m_{2}^{2}} }+x^{2}_{2}]= \nonumber\\
=D^{2}+P^{2}[P^2-4m^{2}_{\tau}][Pp_{1})^{2}/P^{2}-m_{1}^{2}][(Pp_{2})^{2}/P^{2}-m_{2}^{2}]\times \nonumber\\
\times[x_{1}^{2}+2x_{1}x_{2}\cos{\Theta}+x^{2}_{2}]. \nonumber\\
\end{eqnarray}
After substituting $\epsilon^{\alpha\beta\gamma\rho}\epsilon_{\alpha_{1}\beta_{1}\gamma_{1}\rho}P_{\alpha}p_{1\beta}p_{2\gamma}P^{\alpha_{1}}p_{1}^{\beta_{1}}p_{2}^{\gamma_{1}}$ in the form (\ref{eq:eps}) into equation (\ref{eq:main3}), it can be rewritten as a function of the free parameters $x_{1}$, $x_{2}$, and $D^{2}$:

\begin{eqnarray}
\label{eq:mainlimb} 
x_{1}^{2}+2x_{1}x_{2}\cos{\Theta}+x^{2}_{2}=\sin^{2}{\Theta}-\nonumber\\ 
-\frac{\displaystyle D^{2} }{ \displaystyle P^{2}[P^2-4m^{2}_{\tau}][(Pp_{1})^{2}/P^{2}-m_{1}^{2}][(Pp_{2})^{2}/P^{2}-m_{2}^{2}]}.
\end{eqnarray}

We introduce a new dimensionless parameter:

\begin{eqnarray}
\label{eq:A}
\tilde{D}^{2}= \frac{\displaystyle D^{2} }{ \displaystyle P^{2}[(Pp_{1})^{2}/P^{2}-m_{1}^{2}][(Pp_{2})^{2}/P^{2}-m_{2}^{2}][P^{2}-4m_{\tau}^{2}]}.
\end{eqnarray}
The parameter $\tilde{D}^{2}$ is non-negative in the physical domain, since
the denominator of the expression on the right side of (\ref{eq:A}) is positive.
Then of (\ref{eq:mainlimb}) takes the form:

\begin{eqnarray}
\label{eq:mainlimw}
x_{1}^{2}+2x_{1}x_{2}\cos{\Theta}+x^{2}_{2}=\sin^{2}{\Theta}-\tilde{D}^{2}.
\end{eqnarray}

Equation (\ref{eq:mainlimw}) contains only two external parameters:
$\Theta$ and $\tilde{D}$. The choice of $\tilde{D}$ allows us to set the right side to zero,
and then the solution to the left side is $x_{1}=x_{2}=0$. These points correspond to the following expressions:

\begin{eqnarray}
\label{eq:null}
\mu^{2}_{1}(0)=m^{2}_{\tau}+m^{2}_{1}-(Pp_{1}), \nonumber \\
\mu^{2}_{2}(0)=m^{2}_{\tau}+m^{2}_{2}-(Pp_{2}).
\end{eqnarray}
By decreasing the value of the parameter $\tilde{D}^{2}$, we obtain an ellipse in the space $(x_{1},x_{2})$ (Fig. (\ref{fig:ellipse}).
When $\tilde{D}^{2}=0$, the ellipse size becomes maximum, and the equation $\tilde{D}^{2}=0$ is written as:

\begin{figure}[!h]
\centering
\includegraphics[width=.95\textwidth]{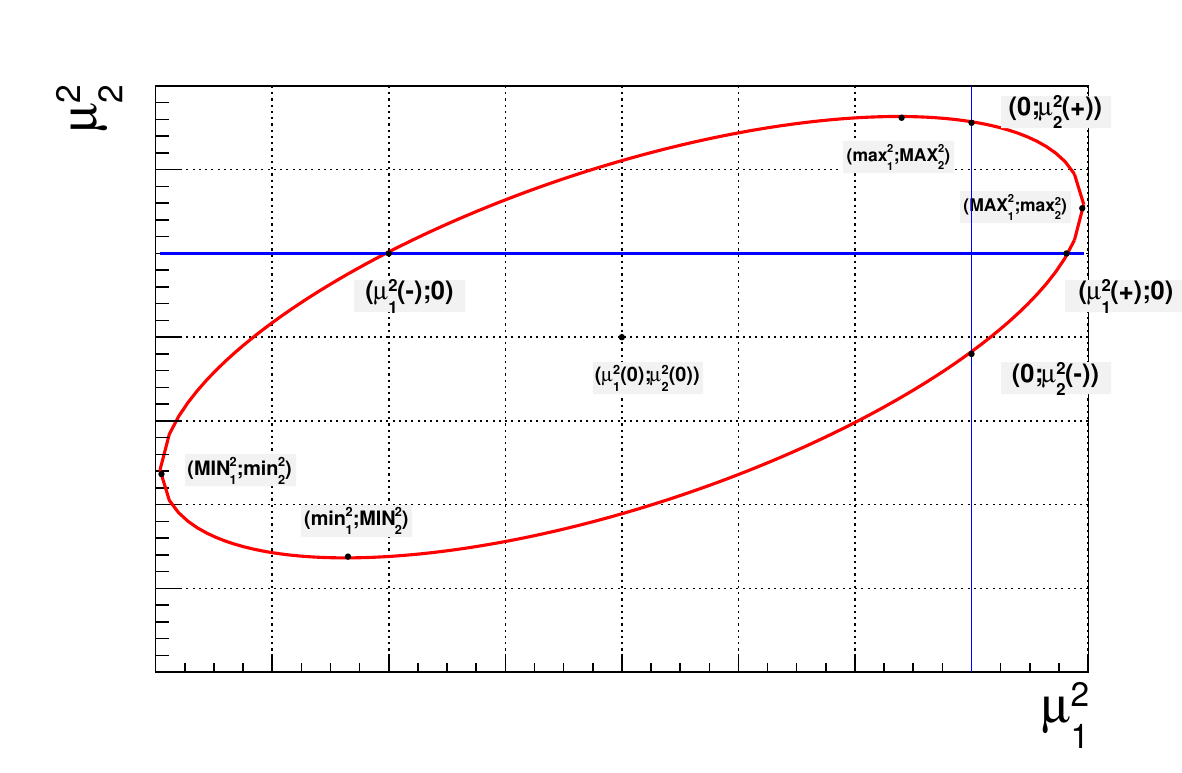}
\caption{\label{fig:ellipse} Ellipse from the equation $D^{2}(\mu^{2}_{1};\mu^{2}_{2})=0$. Blue lines are the coordinate axes. The notations of the absolute maxima and minima,
the position of the center, and the intersection point with the coordinate axes are given.}
\end{figure}

\begin{eqnarray}
\label{eq:mainlimc}
x_{1}^{2}+2x_{1}x_{2}\cos{\Theta}+x^{2}_{2}=\sin^{2}{\Theta}.
\end{eqnarray}
We find $x_{2}$ from the solution of the quadratic equation (\ref{eq:mainlimc})
with respect to $x_{2}$ for fixed $x_{1}$:

\begin{eqnarray}
\label{eq:x1sol}
x_{2}=-x_{1}\cos{\Theta}\pm\sqrt{1-x^{2}_{1}}\sin{\Theta}.
\end{eqnarray}
Note also that the maxima and minima of $x_{i}$ correspond to the maxima and minima of $\mu^{2}_{i}$,
since $\mu^{2}_{i}$ are linear functions of $x_{i}$ and $ \frac{\displaystyle \partial \mu^{2}_{i}}{ \displaystyle \partial x_{i}}\ge0$.
For a quadratic equation to have a solution, the condition:

\begin{eqnarray}
\label{eq:X1}
[P^2-4m^{2}_{\tau}][(Pp_{1})^{2}/P^{2}-m_{1}^{2}]-[m^{2}_{\tau}+m^{2}_{1}-(Pp_{1})]^{2}\geq0,
\end{eqnarray}
must satisfied.
After transformations, we obtain an expression for $\mu^{2}_{2}(+)$ and $\mu^{2}_{2}(-)$:

\begin{eqnarray}
  \label{eq:shiftMU2}
  \mu^{2}_{2}(\pm)=m^{2}_{\tau}+m^{2}_{2}-(Pp_{2})\pm\sqrt{P^2-4m^{2}_{\tau}}\sqrt{(Pp_{2})^{2}/P^{2}-m_{2}^{2}}[-x_{1}\cos{\Theta}+\sqrt{1-x^{2}_{1}}\sin{\Theta}]=\nonumber \\
  =m^{2}_{\tau}+m^{2}_{2}-(Pp_{2})\pm\sqrt{P^2-4m^{2}_{\tau}}\sqrt{(Pp_{2})^{2}/P^{2}-m_{2}^{2}}\times \nonumber \\
  \times\frac{\displaystyle [m^{2}_{\tau}+m^{2}_{1}-(Pp_{1})]\cos{\Theta}+\sqrt{[P^2-4m^{2}_{\tau}][(Pp_{1})^{2}/P^{2}-m_{1}^{2}]-[m^{2}_{\tau}+m^{2}_{1}-(Pp_{1})]^{2}}\sin{\Theta}}{\displaystyle \sqrt{P^2-4m^{2}_{\tau}}\sqrt{(Pp_{1})^{2}/P^{2}-m_{1}^{2}}}=\nonumber \\
  =m^{2}_{\tau}+m^{2}_{2}-(Pp_{2})\pm\frac{\displaystyle 1}{\displaystyle (Pp_{1})^{2}/P^{2}-m_{1}^{2}}\times \nonumber \\
  \times\Big[[m^{2}_{\tau}+m^{2}_{1}-(Pp_{1})][(Pp_{2})(Pp_{1})/P^{2}-(p_{1}p_{2})]+\nonumber \\
    +\sqrt{[(Pp_{2})^{2}/P^{2}-m_{2}^{2}][(Pp_{1})^{2}/P^{2}-m_{1}^{2}]
+[(Pp_{1})(Pp_{2})/P^{2}-(p_{1}p_{2})]^{2}}\times \nonumber \\\sqrt{[P^2-4m^{2}_{\tau}][(Pp_{1})^{2}/P^{2}-m_{1}^{2}]-[m^{2}_{\tau}+m^{2}_{1}-(Pp_{1})]^{2}}\Big].  \nonumber \\
  \end{eqnarray}

When inequality (\ref{eq:Fcond}) is not satisfied, and $\tau^{+}$ decays into leptons (we assume that all particles except neutrinos are registered), then the signal event $\tau^{-}\to\pi^{-}\eta\nu_{\tau}$ satisfies the law of conservation of energy-momentum if:

\begin{eqnarray}
    \label{eq:Z2C}
          [-m^{2}_{\tau}-m^{2}_{1}+(Pp_{1})]^{2}\leq[P^2-4m^{2}_{\tau}][(Pp_{1})^{2}/P^{2}-m_{1}^{2}] \nonumber \\
  \mu^{2}_{2}(+)\geq0  \nonumber \\
\end{eqnarray}

As a result, for the semileptonic decay of the tagging $\tau^{+}$, the fulfillment of
inequality (\ref{eq:Fcond}) is necessary and sufficient, and for the leptonic decay, either
the fulfillment of condition (\ref{eq:Fcond}) or conditions (\ref{eq:Z2C}) is necessary and sufficient.

\subsubsection{ Background from the decay of $\tau^{-}\to\pi^{-}\eta K^{0}\nu_{\tau}$ }
\label{ch:Ks}

Let's find the maximum of $x_{1}$ from equation (\ref{eq:mainlimc}).
First we calculate the derivative $\frac{\displaystyle \partial x_{1}}{ \displaystyle \partial x_{2}}$.

\begin{eqnarray}
\label{eq:div}
\frac{\partial x_{1}}{\partial x_{2}}=-\frac{\displaystyle x_{1}\cos{\Theta} +x_{2} }{ \displaystyle x_{1}+x_{2}\cos{\Theta}}.
\end{eqnarray}
From the condition $\frac{\displaystyle \partial x_{1}}{ \displaystyle\partial x_{2}}=0$, we obtain:

\begin{eqnarray}
\label{eq:x1}
x_{2}=-x_{1}\cos{\Theta}.
\end{eqnarray}
Substitute $x_{2}$ in the form (\ref{eq:x1}) into equation (\ref{eq:mainlimc}):

\begin{eqnarray}
\label{eq:mainlimd}
x_{1}^{2}-2x_{1}^{2}\cos^{2}{\Theta}+x_{1}^{2}\cos^{2}{\Theta}=x^{2}_{1}\sin^{2}{\Theta}=\sin^{2}{\Theta}.
\end{eqnarray}
The absolute maximum point of $x_{1}=1$ along the along the ellipse is reached at the point  $x_{2}=-\cos{\Theta}$.

Let's find the maximum value of $\mu_{1}^{2}$, $MAX_{1}^{2}$, along the ellipse in $\mu_1^2$, $\mu_{2}^2$ plane. This point is shown in Fig.\ref{fig:ellipse} as the point
$(MAX_{1}^{2}, max_{2}^2)$, where $max_{2}^2$ is the value of $\mu_{2}^2$ when $\mu_{1}^2= MAX_{1}^{2}$, substituting the values above into (\ref{eq:shift}):

\begin{eqnarray}
\label{eq:shift1}
MAX^{2}_{1}=m^{2}_{\tau}+m^{2}_{1}-(Pp_{1})+\sqrt{P^2-4m^{2}_{\tau}}\sqrt{(Pp_{1})^{2}/P^{2}-m_{1}^{2}}, \nonumber \\
max^{2}_{2}=m^{2}_{\tau}+m^{2}_{2}-(Pp_{2})-\sqrt{P^2-4m^{2}_{\tau}}\frac{\displaystyle (Pp_{1})(Pp_{2})-(p_{1}p_{2})P^{2} }{\displaystyle \sqrt{(Pp_{1})^{2}-m_{1}^{2}P^{2}}\sqrt{P^{2}} }.
\end{eqnarray}

The absolute minimum $x_{1}=-1$ is reached at the point $x_{2}=\cos{\Theta}$.
We can find absolute minimum $\mu_{1}^{2}$ - $MIN_{1}^{2}$ and the corresponding $\mu_{2}^{2}$ - $min_{2}^{2}$ after  substitution "+" as "-" and "-" as "+" in apropriate parts:

\begin{eqnarray}
\label{eq:shift2}
MIN^{2}_{1}=m^{2}_{\tau}+m^{2}_{1}-(Pp_{1})-\sqrt{P^2-4m^{2}_{\tau}}\sqrt{(Pp_{1})^{2}/P^{2}-m_{1}^{2}}, \nonumber \\
min^{2}_{2}=m^{2}_{\tau}+m^{2}_{2}-(Pp_{2})+\sqrt{P^2-4m^{2}_{\tau}}\frac{\displaystyle (Pp_{1})(Pp_{2})-(p_{1}p_{2})P^{2} }{\displaystyle \sqrt{(Pp_{1})^{2}-m_{1}^{2}P^{2}}\sqrt{P^{2}} }.
\end{eqnarray}

Now consider the background from the decay of $\tau^{-}\to\pi^{-}\eta K^{0}\nu_{\tau}$
with the lost $K^{0}$.
Let $\tau^{+}$ decay semileptonically.
This means that $D^{2}$ from equation (\ref{eq:main3})
takes non-negative values for $\mu^{2}_{1}\geq m^{2}_{K^{0}}$ and
$\mu^{2}_{2}=0$.

We find the maximum of $x_{1}$ from the solution of the quadratic equation (\ref{eq:mainlimc})
with respect to $x_{1}$ for fixed $x_{2}$:

\begin{eqnarray}
\label{eq:x1sol}
x_{1}=-x_{2}\cos{\Theta}\pm\sqrt{1-x^{2}_{2}}\sin{\Theta}.
\end{eqnarray}

For a quadratic equation to have a solution, the condition
$|x_{2}|\leq1$ or:

\begin{eqnarray}
\label{eq:X2}
[P^2-4m^{2}_{\tau}][(Pp_{2})^{2}/P^{2}-m_{2}^{2}]-[m^{2}_{\tau}+m^{2}_{2}-(Pp_{2})]^{2}\geq0.
\end{eqnarray}

By analogy with (\ref{eq:X2}), we denote the maximum and minimum
$\mu^{2}_{1}$ at $\mu_{2}^{2}=0$ as $\mu^{2}_{1}(+)$ and $\mu^{2}_{1}(-)$ (See the points $(\mu_{1}^{2}(+), 0),(\mu_{1}^{2}(-), 0)$ in Fig.\ref{fig:ellipse}).
After transformations, we obtain an expression for the maximum and minimum:

\begin{eqnarray}
\label{eq:shiftMU1}
\mu^{2}_{1}(\pm)=m^{2}_{\tau}+m^{2}_{1}-(Pp_{1})\pm\sqrt{P^2-4m^{2}_{\tau}}\sqrt{(Pp_{1})^{2}/P^{2}-m_{1}^{2}}[-x_{2}\cos{\Theta}+\sqrt{1-x^{2}_{2}}\sin{\Theta}]=\nonumber \\ 
=m^{2}_{\tau}+m^{2}_{1}-(Pp_{1})\pm\sqrt{P^2-4m^{2}_{\tau}}\sqrt{(Pp_{1})^{2}/P^{2}-m_{1}^{2}}\times \nonumber \\ 
\times\frac{\displaystyle [m^{2}_{\tau}+m^{2}_{2}-(Pp_{2})]\cos{\Theta}+\sqrt{[P^2-4m^{2}_{\tau}][(Pp_{2})^{2 }/P^{2}-m_{2}^{2}]-[m^{2}_{\tau}+m^{2}_{2}-(Pp_{2})]^{2}}\sin{\Theta}}{\displaystyle \sqrt{P^2-4m^{2}_{\tau}}\sqrt{(Pp_{2})^{2}/P^{2}-m_{2}^{2}}}=\nonumber \\ 
=m^{2}_{\tau}+m^{2}_{1}-(Pp_{1})\pm\frac{\displaystyle 1}{\displaystyle (Pp_{2})^{2}/P^{2}-m_{2}^{2}}\times \nonumber \\ 
\times\Big[[m^{2}_{\tau}+m^{2}_{2}-(Pp_{2})][(Pp_{1})(Pp_{2})/P^{2}-(p_{1}p_{2})]+\nonumber \\ 
+\sqrt{[(Pp_{1})^{2}/P^{2}-m_{1}^{2}][(Pp_{2})^{2}/P^{2}-m_{2}^{2}]
+[(Pp_{1})(Pp_{2})/P^{2}-(p_{1}p_{2})]^{2}}\times \nonumber \\\sqrt{[P^2-4m^{2}_{\tau}][(Pp_{2})^{2}/P^{2}-m_{2}^{2}]-[m^{2}_{\tau}+m^{2}_{2}-(Pp_{2})]^{2}}\Big]. \nonumber \\
\end{eqnarray}
For a background process, it is necessary and sufficient that:

\begin{eqnarray}
\label{eq:FcondG}
\mu^{2}_{1}(+)\geq m^{2}_{K^{0}} \nonumber \\
\mu^{2}_{1}(-)\leq m^{2}_{K^{0}} \nonumber \\
(P^2-4m^{2}_{\tau})[(Pp_{2})^{2}/P^{2}-m_{2}^{2}]-[m^{2}_{\tau}+m^{2}_{2}-(Pp_{2})]^{2}\geq0.
\end{eqnarray}

If $\tau^{+}$ decays into leptons and conditions (\ref{eq:FcondG})
are not satisfied, an alternative arises
 to achieve positive valuses of $D^{2}$ in the region of $\mu^{2}_{1}\geq m^{2}_{K^{0}}$
and $\mu^{2}_{2}\geq0$. These conditions can be written as:

\begin{eqnarray}
\label{eq:FcondW}
MAX^{2}_{1}\geq m^{2}_{K^{0}}, \nonumber \\
max^{2}_{2}>0.
\end{eqnarray}
Here $MAX^{2}_{1}$ and $max_{2}^{2}$ are taken from equation
(\ref{eq:shift1}). The absolute  maximum value of $\mu_{1}^{2}$ along the ellipse is $max(\mu_{1}^{2})$, subject to the condition that $\mu_{2}^{2}\geq0$
is written as follows.

\begin{eqnarray}
  \label{q:errora}
max(\mu_{1}^{2})=\nonumber \\
=
\begin{cases}
MAX_{1}^{2}, ~if~ max_{2}^{2}\geq0\nonumber \\
\mu^{2}_{1}(+), ~else~if~ (P^2-4m^{2}_{\tau})[(Pp_{2})^{2}/P^{2}-m_{2}^{2}]-[m^{2}_{\tau}+m^{2}_{2}-(Pp_{2})]^{2}\geq0 
\end{cases} \\
\end{eqnarray}

\subsubsection{ Background from the decay of $\tau^{-}\to\rho^{-}\eta\nu_{\tau}$ }

Since $\pi^{0}$ is lost in this process, the conditions will be the same as for the previous background source, with the mass $m^{2}_{K^{0}}$ replaced by $m_{\pi^{0}}^{2}$.
\SKIP{The decay scheme is shown in Fig. (\ref{fig:B}).} Assuming that
the invariant mass of the lost $\pi^{0}$ and $\pi^{-}$ coincides with the 
$\rho^{-}$ meson mass, we can find additional conditions for this background source.
It should be noted that the width of $\rho^{-}$ is quite large, so
in the formulas we obtained, it is more logical to use not the static
mass $M_{\rho^{-}}$, but rather some effective value, the magnitude of which, generally speaking, is determined by the experimental setup.

Let the four-momentum $\pi^{0}$ from the decay $\tau^{-}$ be $p_{\pi^{0}1}$ and the neutrino be $p_{\nu1}$,
and their sum be:

\begin{eqnarray}
p_{\pi^{0}1}^{\mu}+p_{\nu1}^{\mu}= \frac{P^{\mu}+Q^{\mu}}{2}-p_{1}^{\mu}.
\label{eq:11}
\end{eqnarray}
For the decay $\rho^{-}\to\pi^{-}\pi^{0}$ we have the condition:
$(p_{\pi^{-}1}+p_{\pi^{0}1})^{2}=M_{\rho^{-}}^{2}$, where $p_{\pi^{-}1}$ is the four-momentum
$\pi^{-}$ from the decay $\tau^{-}$. To separate the system $\pi^{0}\nu$, we introduce an auxiliary four-vector $q_{1}$ such that

\begin{eqnarray}
p_{\pi^{0}1}^{\mu}=\frac{\displaystyle \mu^{2}_{1}+m^{2}_{\pi^{0}} }{ \displaystyle 2 \mu^{2}_{1}} [p_{\pi^{0}1}^{\mu}+p_{\nu1}^{\mu}]+q^{\mu}_{1},\nonumber \\
p_{\nu1}^{\mu}=\frac{\displaystyle \mu^{2}_{1}-m^{2}_{\pi^{0}} }{ \displaystyle 2 \mu^{2}_{1}} [p_{\pi^{0}1}^{\mu}+p_{\nu1}^{\mu}]-q^{\mu}_{1}, 
\label{eq:12}
\end{eqnarray}
with additional conditions:

\begin{eqnarray} 
(p_{\pi^{0}1}q_{1})+(p_{\nu 1}q_{1})=0, \nonumber \\ 
q_{1}^{2}=\frac{\displaystyle -[\mu^{2}_{1}-m^{2}_{\pi^{0}}]^{2} }{ \displaystyle 4 \mu^{2}_{1}}. 
\label{eq:13}
\end{eqnarray}
The scalar product $(p_{\pi^{0}1}p_{\pi^{-}1})$ is calculated from the condition on the mass $M^{2}_{\rho^{-}}$:

\begin{eqnarray}
(p_{\pi^{0}1}p_{\pi^{-}1})=\frac{\displaystyle M^{2}_{\rho^{-}}-m^{2}_{\pi^{0}}-m^{2}_{\pi^{-}} }{ \displaystyle 2 }=\frac{\displaystyle \mu^{2}_{1}+m^{2}_{\pi^{0}} }{ \displaystyle 2 \mu^{2}_{1}}[(p_{\pi^{0}1}p_{\pi^{-}1})+(p_{\nu1}p_{\pi^{-}1})]+(q_{1}p_{\pi^{-}1}), 
\label{eq:14}
\end{eqnarray}

The kinematic limit can be obtained from the second equality in (\ref{eq:13}).
One coordinate $q_{1}$ in the basis based on the vectors $p_{\pi^{0}1}+p_{\nu 1}, p_{\pi^{-}1}$,
can be obtained from the first equality in (\ref{eq:14}), the second from (\ref{eq:13}), and the third $C_{1}$ is a free parameter,
defined as:

\begin{eqnarray}
q_{1}^{\alpha}(p_{\pi^{0}1}+p_{\nu 1})^{\beta}p_{\pi^{-}1}^{\gamma}\epsilon_{\alpha\beta\gamma}=C_{1}.
\label{eq:15}
\end{eqnarray}

From (\ref{f:metric10}) with substitutions $(u_{1}u_{2})=q^{2}_{1},a=p_{\pi^{0}1}+p_{\nu1},b=p_{\pi^{-}1},A_{1}=A_{2}=0,B_{1}=B_{2}=(q_{1}p_{\pi^{-}1}),C_{2}=C_{1}$; and the variant $q^{2}_{1}$ can be found as:

\begin{eqnarray} 
q_{1}^{2}=\frac{\displaystyle \mu^{2}_{1}[\frac{ M^{2}_{\rho^{-}}-m^{2}_{\pi^{0}}-m^{2}_{\pi^{-}} }{ 2 }-\frac{\displaystyle \mu^{2}_{1}+m^{2}_{\pi^{0}} }{ \displaystyle 2 \mu^{2}_{1}}[(p_{\pi^{0}1}p_{\pi^{-} 1})+(p_{\nu 1}p_{\pi^{-} 1})]]^{2}+C^{2}_{1} }{ \displaystyle \mu^{2}_{1}m_{\pi^{-}}^{2} -[(p_{\pi^{0}1}p_{\pi^{-} 1})+(p_{\nu 1}p_{\pi^{-} 1})]^{2} }.\nonumber \\ 
\label{eq:q21}
\end{eqnarray}
Then $C^{2}_{1}$ is given by the following expression:

\begin{eqnarray} 
4C_{1}^{2}=-\mu^{2}_{1}[ M^{2}_{\rho^{-}}-m^{2}_{\pi^{0}}-m^{2}_{\pi^{-}} -\frac{\displaystyle \mu^{2}_{1}+m^{2}_{\pi^{0}} }{ \displaystyle \mu^{2}_{1}}[(p_{\pi^{0}1}p_{\pi^{-} 1})+(p_{\nu 1}p_{\pi^{-} 1})]^{2}-\nonumber \\ 
-\frac{\displaystyle (\mu^{2}_{1}-m^{2}_{\pi^{0}})^{2} }{ \displaystyle \mu^{2}_{1}}[\mu^{2}_{1}m_{\pi^{-}}^{2} -[(p_{\pi^{0}1}p_{\pi^{-} 1})+(p_{\nu 1}p_{\pi^{-} 1})]^{2}]=\nonumber \\ 
= -4m_{\pi^{0}}^{2}[(p_{\pi^{0}1}p_{\pi^{-} 1})+(p_{\nu 1}p_{\pi^{-} 1})]^{2}+2[M^{2}_{\rho^{-}}-m^{2}_{\pi^{-}}-m^{2}_{\pi^{0}}][(p_{\pi^{0}1}p_{\pi^{-} 1})+(p_{\nu 1}p_{\pi^{-} 1})]\mu_{1}^{2}-\nonumber \\ 
-m_{\pi^{-}}^{2}(\mu_{1}^{2})^{2}+2m^{2}_{\pi^{0}}[M_{\rho^{-}}^{2}-m^{2}_{\pi^{0}}-m_{\pi^{-}}^{2}][(p_{\pi^{0}1}p_{\pi^{-} 1})+(p_{\nu 1}p_{\pi^{-} 1})]-\nonumber \\
-[m^{4}_{\pi^{0}}-2M_{\rho^{-}}^{2}m_{\pi^{0}}^{2}+M^{4}_{\rho^{-}}-2m_{\pi^{0}}^{2}m_{\pi^{-}}^{2}+m_{\pi^{-}}^{4}]\mu^{2}_{1}-m^{4}_{\pi^{0}}m_{\pi^{-}}^{2}.\nonumber \\
\label{f:C1_1}
\end{eqnarray}

To obtain this equation, we multiplied $q^{2}_{1}$ by the denominator of the fraction
on the right side of the equation (\ref{eq:q21}).
Let's consider separately a special case where

\begin{eqnarray}
  \mu^{2}_{1}m_{\pi^{-}}^{2} =[(p_{\pi^{0}1}p_{\pi^{-} 1})+(p_{\nu 1}p_{\pi^{-} 1})]^{2},\nonumber \\
  (q_{1}p_{\pi^{-}1})=(q_{1}p_{\pi^{0}1})+(q_{1}p_{\nu1})=0.
            \label{eq:q22}
\end{eqnarray}

It can be seen that (\ref{eq:14}) implies that
$C_{1}^{2}=0$ when (\ref{eq:q22}) is satisfied.
Equation (\ref{f:C1_1}) is valid,
even if conditions (\ref{eq:q22})
are satisfied.
The kinematic limit can be found from the condition $C^{2}_{1}=0$.
$C^{2}_{1}$ is a quadratic form in the variables $\mu^{2}_{1}$ and $(p_{\pi^{0}1}p_{\pi^{-} 1})+(p_{\nu 1}p_{\pi^{-} 1})$. Let's make a change of variables 
(\ref{f:C1_1}) $x=\mu^{2}_{1}-m_{\pi^{0}}^{2},y=(p_{\pi^{0} 1}p_{\pi^{-} 1})+(p_{\nu 1}p_{\pi^{-} 1})-[M^{2}_{\rho^{-}}-m^{2}_{\pi^{0}}-m^{2}_{\pi^{-}}]/2$ and rewrite (\ref{f:C1_1})

\begin{eqnarray} 
4C_{1}^{2}(x;y)=-4m^{2}_{\pi^{0}}y^{2}-m^{2}_{\pi^{-}}x^{2}+2[M^{2}_{\rho^{-}}-m^{2}_{\pi^{0}}-m^{2}_{\pi^{-}}]xy=\nonumber \\ 
=-4m^{2}_{\pi^{0}}[y-\frac{[-m^{2}_{\pi^{0}}+M^{2}_{\rho^{-}}-m^{2}_{\pi^{-}}]x-\sqrt{[[M^{2}_{\rho^{-} }-m^{2}_{\pi^{0}}-m^{2}_{\pi^{-}}]^{2}-4m^{2}_{\pi^{0}}m^{2}_{\pi^{-}}]x^{2}}}{4m^{2}_{\pi^{0}}}]\times \nonumber\\
\times[y-\frac{[-m^{2}_{\pi^{0}}+M^{2}_{\rho^{-}}-m^{2}_{\pi^{-}}]x+\sqrt{[[M^{2}_{\rho^{-}}-m^{2}_{\pi^{0}}-m^{2}_{\pi^{-}}]^{2}-4m^{2}_{\pi^{0}}m^{2}_{\pi^{-}}]x^{2}}}{4m^{2}_{\pi^{0}}}].\nonumber \\
\label{f:C1_2}
\end{eqnarray}

The equation $C^{2}_{1}(x;y)=0$ yields two lines intersecting at the origin.
The situation is similar to the example from paragraph (\ref{sec:pm}) equation (\ref{f:cone}) figure (\ref{fig:sector}). This is the light cone describing the motion of a neutrino with zero mass.
In the physical region $x\geq0$, and the condition $C_{1}^{2}\geq0$ yields two inequalities:

\begin{eqnarray}
y\geq\frac{-m^{2}_{\pi^{0}}+M^{2}_{\rho^{-}}-m^{2}_{\pi^{-}}-\sqrt{[M^{2}_{\rho^{-}}-m^{2}_{\pi^{0}}-m^{2}_{\pi^{-}}]^{2}-4m_{\pi^{0}}^{2}m^{2}_{\pi^{-}}}}{4m^{2}_{\pi^{0}}}x, \nonumber \\
y\leq\frac{-m^{2}_{\pi^{0}}+M^{2}_{\rho^{-}}-m^{2}_{\pi^{-}}+\sqrt{[M^{2}_{\rho^{-}}-m^{2}_{\pi^{0}}-m^{2}_{\pi^{-}}]^{2}-4m^{2}_{\pi^{0}}m^{2}_{\pi^{-}}}}{4m^{2}_{\pi^{0}}}x.
\label{f:C1_3}
\end{eqnarray}
This is a sector in the $(x;y)$ plane. The entire plane is divided into four sectors by equation (\ref{f:C1_2}) with the condition $C_{1}=0$, but
only one of them represents the physical region.
Now we can express $x$ and $y$ as a function of $Q$ from equation (\ref{eq:11}):

\begin{eqnarray}
x=m^{2}_{\tau}-(Pp_{1})-(Qp_{1})+m_{1}^{2}-m_{\pi^{0}}^{2}, \nonumber \\
y=(Pp_{\pi^{-}1})/2+(Qp_{\pi^{-}1})/2-(p_{1}p_{\pi^{-}1})-[M^{2}_{\rho^{-}}-m^{2}_{\pi^{0}}-m^{2}_{\pi^{-}}]/2,\nonumber \\
\label{f:xy}
\end{eqnarray}
and substitute into the equations (\ref{f:C1_3}).

\begin{eqnarray} 
(Pp_{\pi^{-}1})/2+(Qp_{\pi^{-}1})/2-(p_{1}p_{\pi^{-}1})-[M^{2}_{\rho^{-}}-m^{2}_{\pi^{0}}-m^{2}_{\pi^{-}}]/2\geq\nonumber \\ 
\geq\frac{-m^{2}_{\pi^{0}}+M^{2}_{\rho^{-}}-m^{2}_{\pi^{-}}-\sqrt{[M^{2}_{\rho^{-}}-m^{2}_{\pi^{0}}-m^{2}_{\pi^{-}} ]^{2}-4m_{\pi^{0}}^{2}m^{2}_{\pi^{-}}}}{4m^{2}_{\pi^{0}}}[M^{2}_{\tau}-(Pp_{1})-(Qp_{1})+m_{1}^{2}-m_{\pi^{0}}^{2}], \nonumber\\ 
(Pp_{\pi^{-}1})/2+(Qp_{\pi^{-}1})/2-(p_{1}p_{\pi^{-}1})-[M^{2}_{\rho^{-}}-m^{2}_{\pi^{0}}-m^{2}_{\pi^{-}}]/2\leq\nonumber \\ 
\leq\frac{-m^{2}_{\pi^{0}}+M^{2}_{\rho^{-}}-m^{2}_{\pi^{-}}+\sqrt{[M^{2}_{\rho^{-}}-m^{2}_{\pi^{0}}-m^{2}_{\pi^{-}}]^{2 }-4m^{2}_{\pi^{0}}m^{2}_{\pi^{-}}}}{4m^{2}_{\pi^{0}}}[M^{2}_{\tau}-Pp_{1}-(Qp_{1})+m_{1}^{2}-m_{\pi^{0}}^{2}].\nonumber \\ 
\label{f:C1_4}
\end{eqnarray}
From Eqs. (\ref{f:QCon}) we obtain:

\begin{eqnarray}
(PQ)=0,\nonumber \\
Q^{2}=4m^{2}_{\tau}-P^{2}.
\label{f:QCon1}
\end{eqnarray}

For the decay event $\tau^{-}$ to be compatible with
the hypothesis that $\tau^{-}\to\rho^{-}\eta\nu_{\tau}\to\pi^{-}\pi^{0}\eta\nu_{\tau}$ decayed,
and $\pi^{0}$ was not reconstructed, it is necessary that there exists a four-vector
$Q$ satisfying the equations (\ref{f:QCon1}) and the inequalities (\ref{f:C1_4}).

Thus, by testing each event for compatibility of its kinematics with the hypotheses
that the event is a signal, a background event with a lost $K^{0}_{S}$, or a background event with a lost $\pi^{0}$, it is possible to find events that satisfy only the first hypothesis. This will significantly reduce the background level, generally with a loss of efficiency,
but may increase the overall sensitivity.

\section{Initial and final state radiation}

A radiated photon  that is not detected can
easily be accounted for in the equations. Its accounting in the balance depends
on the system emitting it. In this description, we omit effects related to diagram interference,
where photons are emitted from different lines of Feynman diagrams.
As an example, we examine the process presented
in the section \ref{ch:Ks}.

In the first case, we assume that photon is emitted before
$\tau$ pair decays. Instead $e^{+}e^{-}\to\tau^{+}\tau^{-}$
process we have $e^{+}e^{-}\to\tau^{+}\tau^{-}\gamma$.
Let us denote the 4-momentum of the undetected photon as
$k$.
In all the equations we obtained in Section 3,
we must perform the following substitution
$P\to P-k$. After this, they will correctly account for the emitted photon.
We have three new parameters
$(kP),(kp_{1}),(kp_{2})$
 the scalar products of the photon four-momentum with our basis four-vectors.
The fourth coordinate is:
\begin{eqnarray}
  D_{\rm{rad}}=\epsilon_{\alpha\beta\gamma\rho}k^{\alpha}P^{\beta}p_{1}^{\gamma}p_{2}^{\rho}
\label{f:photon}
\end{eqnarray}
Also we have additional condition on
these parameters:
\begin{eqnarray}
  D^{2}_{\rm{rad}}\geq0
  \label{f:Dphoton}
\end{eqnarray}

During its calculation  we have to put $k^{2}=0$.
Furthermore, care must be taken to ensure that particles
do not transition from the initial state to the final state, or vice versa ($(kP)\geq0$).
$ D^{2}_{\rm{rad}}$ as  function of $(kP),(kp_{1}),(kp_{2})$ is a quadratic form with an inertia index of -1,
one positive and two negative  matrix eigenvalues.

In the second case photon emission in $\tau$-lepton decays.
Namely occurs $\tau^{+}\to l^{+}\bar{\nu_{\tau}}\nu_{l}\gamma$
or $\tau^{+}\to h^{+}\bar{\nu_{\tau}}\gamma$ for tagging
side and $\tau^{-}\to \pi^{-}\eta\bar{\nu_{\tau}}\gamma$
for signal side (here $h^{+}$ is some hadron state from $\tau^{+}$
with four-momentum $p_{2}$). 

In case photon emission in tag $\tau$-lepton decay,
the effect depends on tag decay mode. 
For the process $\tau^{+}\to l^{+}\bar{\nu_{\tau}}\nu_{l}\gamma$  photon emission
don't change the limits of $\mu^{2}_{2}$, it remains non-negative.
This does not reduce the effectiveness of the suppression.

For the process $\tau^{+}\to h^{+}\bar{\nu_{\tau}}\gamma$ the limit on parameter
$\mu^{2}_{2}$ becomes $\mu^{2}_{2}\geq0$ instead of $\mu^{2}_{2}=0$.
The same condition as it has in the lepton tag case.
Such radiation reduces efficiency. 

The photon emission in the signal process $\tau^{-}\to \pi^{-}\eta\gamma\nu_{\tau}$ also changes the limit on $\mu^{2}_{1}$, the same way
as for the hadron tag case. Instead of $\mu^{2}_{1}=0$,
the limit become $\mu^{2}_{1}\geq0$.

In general, photon emission smears the kinematic
constraints derived in its absence,
complicating the isolation of signal processes and
the suppression of background.

\section{Conclusion}

A method for finding kinematic
limits is presented using a coordinate system
based on the four-momenta of the particles participating in the reaction.
Such coordinate systems allow to easily
take into account four-momentum conditions in a form
widely used in particle physics.
A general recipe for obtaining sign-definite
invariants is demonstrated. Kinematic limits are realized when
these invariants are zero. Note that these invariants are analogous to the Cayley-Menger determinants for Minkowski space.
Several examples of using the proposed
approach to analyze physical processes are given.

The use of this method can improve
sensitivity, increase measurement accuracy, and
reduce background for processes with incomplete reconstruction.
This is because it allows to use all available information
on the kinematics of an event and the formation of necessary and sufficient criteria
for an event satisfying the kinematic hypothesis in the simplest possible form. It can also be used to select events for calibration purposes.

\acknowledgments

When using this method to study the properties of the $\tau$-lepton decays with
 the Belle experiment, it was actively discussed within the "tau-2photon"  working group of the Belle detector. I wanted to highlight the many years of joint work with Hisaki Hayashii and Denis Epifanov. In preparing
this text, decisive contributions to its quality were made by
Alexey Garmash and Anna Vinokurova.


\end{document}